\documentclass[fleqn,usenatbib]{mnras}

\usepackage{newtxtext,newtxmath}

\usepackage[T1]{fontenc}
\usepackage{ae,aecompl}

\usepackage{graphicx,subfigure}	
\usepackage{amsmath}	
\usepackage{amssymb}	
\usepackage{hyperref}   
\usepackage{booktabs}
\usepackage[dvipsnames]{xcolor}
\usepackage[utf8]{inputenc} 

\title[Yields from Neutron Star Accretion]{Nucleosynthetic Yields from Neutron Stars Accreting in Binary Common Envelopes}

\author[Keegans, J. et al.]{
J. Keegans$^{1,8}$\thanks{E-mail: j.keegans@2016.hull.ac.uk (KTS)},
C.~L.~Fryer$^{2,3,4,5,8}$,
S.~W.~Jones$^{2,8}$,
B.~C\^ot\'e$^{8,9,10}$,
K.~Belczynski$^{6}$,
\newauthor
F.~Herwig$^{7,8,9}$,
M.~Pignatari$^{1,8,9,10}$,
A.~M.~Laird$^{11,8}$
and
C. Aa. Diget$^{11}$
\\
$^{1}$E. A. Milne Center for Astrophysics, University of Hull, Hull, HU6 7RX, UK \\
$^{2}$Computational Physics and Methods Group (CCS-2), Los Alamos National Laboratory, New Mexico 87544, USA \\
$^{3}$Department of Astronomy, The University of Arizona, Tucson, AZ 85721, USA \\
$^{4}$Department of Physics and Astronomy, The University of New Mexico, Albuquerque, NM 87131, USA \\
$^{5}$The George Washington University, Washington, DC 20052, USA \\
$^{6}$Nicolaus Copernicus Astronomical Center, Polish Academy of Sciences, ul. Bartycka 18, 00-716 Warsaw, Poland \\
$^{7}$Department of Physics \& Astronomy, University of Victoria, Victoria, BC,V8P5C2 Canada.\\
$^{8}$NuGrid Collaboration, http://nugridstars.org\\
$^{9}$Joint Institute for Nuclear Astrophysics - Center for the Evolution of the Elements, USA\\
$^{10}$Konkoly Observatory, Research Center for Astronomy and Earth Sciences, Hungarian Academy
of Sciences, \\ Konkoly Thege Miklos ut 15-17, H-1121 Budapest, Hungary.\\
$^{11}$Department of Physics, University of York, Heslington, York YO10 5DD, U.K\\
}

\date{Accepted XXX. Received YYY; in original form ZZZ}

\pubyear{2015}

\begin{document}
\label{firstpage}
\pagerange{\pageref{firstpage}--\pageref{lastpage}}
\maketitle

\begin{abstract}

Massive-star binaries can undergo a phase where one of the two stars expands during its advanced evolutionary stage as a giant and envelops its companion, ejecting the hydrogen envelope and tightening its orbit. Such a common envelope phase is required to tighten the binary orbit in the formation of many of the observed X-ray binaries and merging compact binary systems. In the formation scenario for neutron star binaries, the system might pass through a phase where a neutron star spirals into the envelope of its giant star companion. These phases lead to mass accretion onto the neutron star.  Accretion onto these common-envelope-phase neutron stars can eject matter that has undergone burning  near to the neutron star surface.  This paper presents nucleosynthetic yields of this ejected matter, using population synthesis models to study the importance of these nucleosynthetic yields in a galactic chemical evolution context.  Depending on the extreme conditions in temperature and density found in the accreted material, both proton-rich and neutron-rich nucleosynthesis can be obtained, with efficient production of neutron rich isotopes of low Z material at the most extreme conditions, and  proton rich isotopes, again at low Z, in lower density models. 
Final yields are found to be extremely sensitive to the physical modeling of the accretion phase. We show that neutron stars accreting in binary common envelopes might be a new relevant site for galactic chemical evolution, and therefore more comprehensive studies are needed to better constrain nucleosynthesis in these objects. 



\end{abstract}




\section{Introduction}

A number of dedicated observing programs have shown that most massive stars are formed in systems with at least one binary companion~\citep{2007ApJ...670..747K,2007ApJ...664.1102K,2009AJ....137.4608K,2012ApJ...751....4K,2012Sci...337..444S,2014ApJS..213...34K}.  Those binaries in tight orbits can undergo one or more mass transfer phases where mass from a star, typically as it expands (e.g. in a giant phase) and overfills its Roche lobe, flows onto its companion.  If the expansion is faster than the companion can incorporate the overflowing mass, the system can go through a common envelope (CE) phase where the expanding star envelops its companion, causing the core of the expanding star and its companion to share a common envelope.  The CE phase causes the binary to tighten its orbit and is postulated to explain many of the tight-orbit, massive-star binaries~\citep{2013A&ARv..21...59I}.

In the formation of a variety of massive star binaries including X-ray binaries and double
compact object systems, the stellar system  evolves through one, and often two, CE phases.  In
the first CE phase, the more massive star (primary) evolves off the main sequence, enveloping
its companion.  In some cases, the resultant tightening of the orbit produces a binary that is
sufficiently close that, even after the subsequent collapse and explosion of the primary, the
wind of the companion can accrete onto the neutron star.   This is a common scenario behind the
production of massive X-ray binaries. In some cases, when the companion star evolves off the main sequence, a second CE phase where a neutron star is enveloped by the companion occurs.  This can tighten the orbit  prior to the supernova explosion of the companion that produces binary pulsar systems and compact binaries that are believed to be the site of short-duration gamma-ray bursts \citep[GRBs;][]{1999ApJ...526..152F,1999MNRAS.305..763B}, including merging neutron star systems~\citep{2012ApJ...759...52D,2013ApJ...779...72D,2015ApJ...806..263D} like the one recently detected by advanced LIGO/Virgo~\citep{2017ApJ...848L..12A}.  If the neutron star (NS) spirals into the core of the companion, it can produce a long-duration GRB, the so-called helium-merger model~\citep{1998ApJ...502L...9F}.

The ultimate fate of the binary in this CE phase depends on the masses of the stars and the orbital separation at the onset of the phase.  Many systems eject the hydrogen envelope, forming a binary consisting of a helium star and a NS.  Others do not have sufficient orbital energy to eject the hydrogen envelope prior to merging with the helium core.  These helium mergers were initially proposed to be a long-lived giant star powered by a central Eddington-rate accreting neutron star known as a Thorne-Zytkow object~\citep{1975ApJ...199L..19T}.  However, calculations including neutrino processes found that most of the energy released in the neutron star accretion would be radiated efficiently through neutrinos, allowing the neutron star to accrete at the Bondi-Hoyle rate, causing it to rapidly collapse to a black hole~\citep{1996ApJ...460..801F,1998ApJ...502L...9F}.  

This helium-merger system, forming a black hole accreting system became one of the proposed black hole accretion disk (BHAD) GRB models~\citep{1999ApJ...518..356P,2001ApJ...550..357Z}.  Subsequent simulations have studied the potential of this system to produce ultra-long duration gamma-ray bursts or peculiar supernovae~\citep{2013ApJ...764..181F,2018MNRAS.475.1198S}.  

Material accreting onto neutron stars is not completely incorporated into the neutron star. If the material has enough angular momentum, it can form a disk that could ultimately drive a jet.  This is believed to be rare in most CE scenarios~\citep{2017ApJ...845..173M}.  For neutron star systems, even if the material does not have a sufficient angular momentum to form a disk, some of the accreting material will be reheated and ejected~\citep{2006ApJ...646L.131F, 2009ApJ...699..409F}.  During this accretion process, temperatures and densities become so high that both neutrino emission (that can alter the electron fraction) and nuclear burning can significantly alter the composition of the material.  For the high accretion rates of supernova fallback, the reheated ejecta can burn into heavy r-process elements~\citep{2006ApJ...646L.131F}.  Fallback accretion rates range from a few times $10^{-3}$ to $1~ \, {\rm M_{\odot} s^{-1}}$.  CE accretion rates are typically lower than these rates:  ranging from $10^{-4}-10^6~ \,\rm{{M}_{\odot}yr^{-1}}$ (note the former is in per second while the latter is per year). 

In this paper, we will study the yields from these lower CE accretion rates. In section~\ref{sec:mdot} we review NS accretion in CE, estimating accretion rates for a range of stellar models at different phases in the star's life.  In section~\ref{sec:yields} we review the range of yields expected as a function of accretion rate from our single zone
models.  To determine the effect CE yields have on galactic chemical evolution, we must calculate the distribution of binaries and CE scenarios.  By using these distributions and stellar models, we can estimate the accretion rates.  In  section~\ref{sec:binary}, we use binary population systems to study yields from stars and stellar populations.  We conclude in section \ref{sec:conclusions} with a discussion of the role these yields play in galactic chemical evolution.

\section{NS Accretion in CE Evolution}
\label{sec:mdot}

\subsection{Estimating Mass Accretion}
\label{sec:mass_accretion_est}

When a massive-star companion in a binary with a NS overfills its Roche lobe, its material accretes onto the neutron star.  The accretion rate can be much faster than the NS can incorporate, ultimately developing into a CE phase.  A number of assumptions are made in estimating this accretion rate and, especially for NSs, there seems to be some confusion on the validity of these assumptions.  Here we review the basic physics assumptions and approximations used in this paper to estimate accretion rates.  In astrophysics, the standard estimate for accretion onto a point source is the Bondi-Hoyle-Littleton solution \citep{1941MNRAS.101..227H,1952MNRAS.112..195B}.  The Bondi radius ($R_{\rm B}$) for a neutron star of mass ($M_{\rm NS}$) can be determined by the radius that material of velocity $v$ is bound to the NS, i.e.: 
\begin{equation}
v^2/R_{\rm B} = G M_{\rm NS}/R_{\rm B}^2 \rightarrow R_{\rm B} = G M_{\rm NS}/v^2
\end{equation}
where $G$ is the gravitational constant.  In the simplest case, $v$ is set to the sound speed ($c_s$).  But if the NS is moving with respect to the material, the relative motion ($v_m$) should also be included.  One simple, and often standard, way to include both velocities is through a quadratic sum: $v=\sqrt{v_m^2+c_s^2}$:
\begin{equation}
R_{\rm B} = 2 G M_{\rm NS}/(c_s^2+v_m^2)
\end{equation}
The accretion rate is roughly the mass within this Bondi radius divided by the free-fall time at this radius.  More accurately, this accretion rate is:
\begin{equation}
\dot{M}_{\rm B}=\lambda_{\rm BHL} 4 \pi R_{\rm B}^2 \rho v
\label{eq_acc_rate}
\end{equation}
where $\rho$ is the density of the ambient medium and $\lambda_{\rm BHL}$ is a non-dimensional parameter:  $\lambda_{\rm BHL}=2/(3\pi)$ if we assume free-fall.  Calculations of Bondi-Hoyle accretion allow 
refinement of the value for $\lambda_{\rm BHL}$ and 
determination of the accuracy of our solution to include the different velocity terms~\citep{1994A&AS..106..505R}.  In scenarios like our CE phase, there is both a velocity and density gradient across the Bondi radius, and these features drive instabilities in the accretion that can decrease the accretion rate~\citep{1994ApJ...427..351R,1994ApJ...427..342R,macleod2014accretion,2015APS..APR.U2004M,2017ApJ...838...56M,2017ApJ...845..173M}.

Bondi accretion also assumes that matter falling onto the neutron star accretes passively onto the neutron star.  However, the gravitational potential energy released as matter accretes onto the neutron star is emitted in radiation and matter outflows that can significantly decrease the accretion rate below the Bondi rate.  The Eddington limit is an extreme case of this radiative feedback that assumes all of the energy released is converted into radiation, and the momentum carried by this radiation exerts a force on the inflowing material.  This radiation limits the amount of accretion onto an object.  The radiative force at radius $r$ is:
\begin{equation}
F_{\rm rad} = \frac{L_{\rm rad}}{4 \pi r^2} \frac{\sigma}{c}
\end{equation}
where $L_{\rm rad}$ is the radiative luminosity, $\sigma$ is the cross section, and $c$ is the speed of light.  Setting this force equal to the gravitational force ($F_{\rm grav}=G M_{\rm NS} m/r^2$ where $m$ is the mass of the accreting particle), we derive the Eddington luminosity:
\begin{equation}
L_{\rm Edd} = 4 \pi G M_{\rm NS} m/ (\sigma c).
\end{equation}
For accreting, fully ionized hydrogen, $m$ is the proton mass and $\sigma$ can be set to the Thompson cross-section.  Although most studies use these assumptions to calculate the Eddington luminosity, for some scenarios, such as fallback accretion in supernova, the opacity per unit mass can be much higher~\citep{1999ApJ...511..885F}, lowering the Eddington accretion rate.

The Eddington limit on accretion assumes that the radiative luminosity is equal to the gravitational potential energy released:
\begin{equation}
\dot{M}_{\rm Edd} = 4 \pi m r_{\rm NS}/(\sigma c) 
\end{equation}
where $r_{\rm NS}$ is the neutron star radius.  This limit on the accretion rate assumes spherical symmetry, the radiation is not trapped in the flow and that all the accretion energy is released in radiation.  For low accretion rates, many of these assumptions are valid. But as the accretion rate increases, these assumptions lose their validity.  

Here we review each assumption individually.  Determining whether the radiation is trapped in the flow is difficult without full calculations, but a first order estimate can be made by comparing the diffusive transport velocity ($v_{\rm diff}$):
\begin{equation}
v_{\rm diff} = \lambda/D c
\end{equation}
where $D$ is the size of the transport region (some fraction of the stellar radius) to the infall velocity, typically set to the free-fall velocity ($v_{\rm ff}$):
\begin{equation}
v_{\rm ff} = \sqrt{2G M_{\rm enc}/r}
\end{equation}
where $M_{\rm enc}$ is the enclosed mass of the star at radius $r$.  Using these approximations, 
it is found that, except at the beginning of the CE phase when the neutron star is in the outer layers of the hydrogen envelope when $\dot{M} < 10^{-4} ~\rm{{M}_{\odot}yr^{-1}}$, the radiation is trapped in the flow~\citep{1993ApJ...411L..33C,1996ApJ...460..801F}.  Recall that, for our nucleosynthesis models, we are only concerned with accretion rates above 1 $\rm{{M}_{\odot}yr^{-1}}$ where the radiation is truly trapped in the flow ($v_{\rm diffusion} << v_{\rm infall}$) and the assumptions needed for the Eddington limit are not applicable.

What about the assumption that all of the energy is emitted in photons?  If the radiation is trapped in the flow, the material will shock and settle onto the neutron star.  By calculating the post-shock entropy of the material and assuming it piles onto the neutron star, we can derive the temperature and density properties of this accreted material.  These estimates find that, unless the entropy is above $10,000 {~\rm k_B \, per \, nucleon}$, the temperature and density conditions are such that most of the gravitational energy released will be converted into neutrinos and escape the star without impeding the inflow~\citep{1989ApJ...346..847C,1996ApJ...460..801F}.  In CE models, the entropy ranges between $10-100 {~\rm k_B \, per \, nucleon}$ and, to date, no one has constructed a way to avoid rapid neutrino cooling in CE scenarios.  For CE systems, only a fraction of the energy is emitted in photons and this assumption of the Eddington approximation is also never satisfied.  For systems where the CE phase ends up with the NS merging with its companion's helium core, neutrino emission increases dramatically, allowing the neutron star to incorporate material at the high Bondi rates predicted for these dense conditions.  This merger ultimately forms a rapidly accreting black hole that may produce ultra-long gamma-ray bursts~\citep{1998ApJ...502L...9F}.

Although it seems that Bondi-Hoyle accretion assumptions are most applicable to our problem, not all of the energy is converted to neutrinos and photons, some goes into kinetic energy that ejects a fraction of the accreting material (see Section~\ref{sec:ejecta}).  It is this ejecta that is the subject of our nucleosynthesis study.  We will discuss this ejecta in more detail in Section~\ref{sec:ejecta}, but it is important to understand that the ejecta may also alter the accretion.  We will decrease the accretion rate in our Bondi-Hoyle solution to approximate this effect.  In this project we use two bounds to match the range of efficiencies in simulations: $\lambda_{\rm BHL}=1/4, 1/40$.  The lower value is set to try to capture both asymmetric accretion and ejecta affects that lower the accretion rate.  

\begin{figure}
\centering
\caption{Radius evolution of the stellar models used in this study. The x-axis is time remaining until core collapse (or, in the case of some of the 8~$\rm{M_\odot}$ models, envelope ejection). The points indicate at which times stellar structures were used from the models in order to calculate the accretion rates.}
\label{fig:stellarmodels}
\includegraphics[width=\columnwidth]{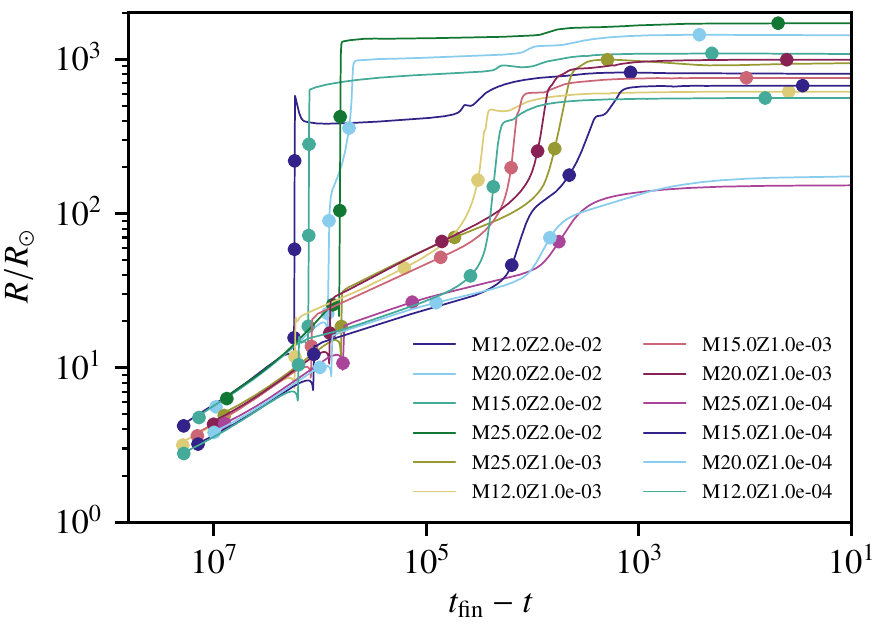}
\includegraphics[width=\columnwidth]{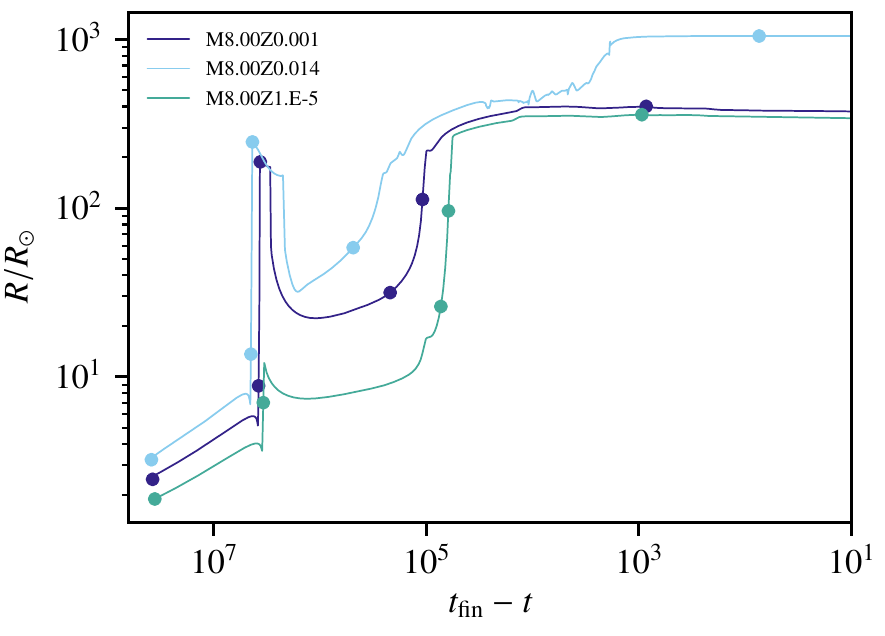}
\end{figure}

The accretion during the CE phase depends on the structure of the star which, in turn, depends upon both the stellar mass and evolutionary stage.  To estimate the NS accretion rates in the CE phase, we use a coarse grid of stellar models computed using the MESA stellar evolution code~\citep{Paxton2011a,Paxton2013a,Paxton2015a,Paxton2018a}, ranging from 8 to 25 solar masses with initial metallicities in the range $10^{-4}\leq Z \leq 2\times10^{-2}$. The massive star models ($M_\mathrm{ini} \geq 12~\rm{M_\odot}$) are the ones from \citet{ritter2018nugrid}. The $8~\rm{M_\odot}$ models were computed using the same input physics as in \citet{Jones2013a}. We refer the reader to those papers for a more thorough description of the stellar evolution calculations.

\begin{figure}
	\includegraphics[width=\columnwidth,clip=true,trim=0cm 3.5cm 0cm 4cm]{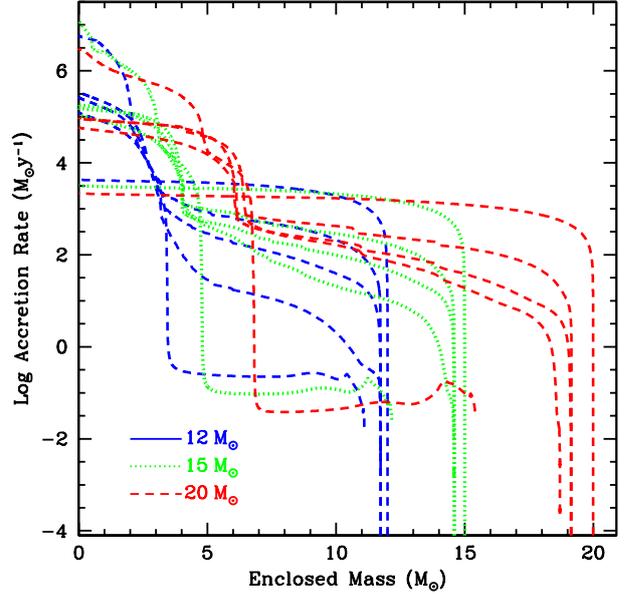}
    \caption{Accretion rate for a 1.4\,M$_\odot$ neutron star as a function of the position of the neutron star within the mass coordinate of its companion for the 12, 15 and 20\,M$_\odot$, solar metallicity stars for a range of times.  At early times, the envelope is compact and the accretion rate is higher.  As the star expands, the accretion rate in the envelope decreases.  In the bulk of the envelope, the accretion rate lies between $1-10^4 {\rm \, M_\odot \, yr^{-1}}$.  Here we assume $\lambda_{\rm BHL}=1/4$.  The accretion rates may be an order of magnitude lower.}
    \label{fig:mdot}
\end{figure}

We study the structure of each of these models as the star evolves, focusing on periods of time when the star is expanding and a CE phase is likely to occur. Figure~\ref{fig:stellarmodels} shows the radius evolution of our stars as a function of time with the points showing the specific times used in our study.  As the neutron star spirals into the star, the accretion rate onto it increases.  The corresponding accretion rates (assuming $\lambda_{\rm BHL}=1/4$) as a function of radius for the 12, 15 and 20\,M$_\odot$ stars at different  evolutionary times is shown in Figure~\ref{fig:mdot}.  In the bulk of the envelope, the accretion rate lies between $1-10^5 {\rm \, M_\odot \, yr^{-1}}$ and we will focus on these rates, but if the neutron star spirals into the core, the accretion rate will be higher. 

\subsection{Ejection of Accreted Mass}
\label{sec:ejecta}

The accreted material is explosively unstable and early calculations suggested that some of the infalling material would gain enough energy to be ejected~\citep{1996ApJ...460..801F}.  Estimates of the convective timescale using the Brunt-V\"ais\"ala frequency~\citep[see, for example,][]{1983apum.conf.....C,1996ApJ...460..801F} suggest that the timescale that the material spends near the proto-neutron star surface is milliseconds in duration.  This initial study focused on CE accretion scenarios, but most of the subsequent, more systematic, multi-dimensional work focused on the higher accretion rates seen in supernova fallback~\citep{2006ApJ...646L.131F,2009ApJ...699..409F}.  Although these studies focused on accretion rates above $10^{4}$\,M$_\odot$\,$yr^{-1}$, they showed the same features as the CE models studied in \cite{1996ApJ...460..801F}.  The accreted material falls down toward the proto-neutron star surface.  A fraction of this material is heated and accelerated to above escape velocities, ejecting it from the system.  

We designed a set of twelve trajectories based on these simulations, guided by the analytic models developed to understand these simulations. The accreting material accelerates nearly at free-fall until it falls within 10\,km of the neutron star surface.  The uncertainty in the flow lies in determining how quickly the flow is reversed and material is ejected.  We study two extremes:  a bounce scenario where the reverse is instantaneous, a convective scenario where the acceleration timescale is on par with the convective turnover timescale.  For the latter, convective timescale, we can 
estimate the acceleration timescale from the Brunt=V\"ais\"ala freqency ($t_{\rm acceleration} = 1/\omega_{\rm BV}$) where
\begin{equation}
\omega_{\rm BV}^2 = g/\rho (\partial \rho/\partial S)_P (\partial S/\partial r),
\end{equation}
where $\rho$ and $S$ are the density and entropy respectively\citep{cox83}.  For conditions in supernovae and fallback this timescale is on the range of 2\,ms\citep{fryer07}.  This timescale estimates the growth timescale of Rayleigh-Taylor instabilities, but this provides an approximate timescale for the acceleration timescale to reverse the shock. Fallback simulations suggest that the true answer lies between these two extremes.  

We model these two extremes by two parameterized simulations:  bounce and convective trajectories.  The bounce trajectory assumes that material falls in adiabatically at free-fall until it reaches 20\,km (10\,km from the surface) and where it is assumed that the material bounce and is ejected at the escape velocity.  For the convective trajectory, we assume that the material falls in at free-fall until it reaches a depth of 50\,km, where we turn on a force that is strong enough to turn around the trajectory within the roughly 2\,ms convective timescale.

\begin{figure}
\includegraphics[width=0.5\textwidth,clip=true,trim=0cm 3.5cm 0cm 4cm]{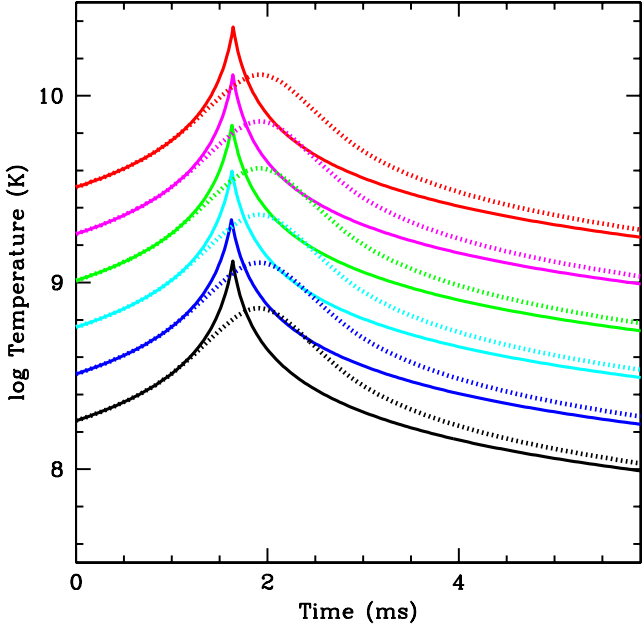} 
\caption{Temperature as a function of time for our trajectories at different accretion rates: 1$~\rm{{M}_{\odot}yr^{-1}}$ (black lines), 10$~\rm{{M}_{\odot}yr^{-1}}$ (blue lines), 10$^{2}~\rm{{M}_{\odot}yr^{-1}}$ (cyan lines), 10$^{3}~\rm{{M}_{\odot}yr^{-1}}$ (green lines), 10$^{4}~\rm{{M}_{\odot}yr^{-1}}$ (magenta lines) and 10$^{5}~\rm{{M}_{\odot}yr^{-1}}$ (red lines).  Solid lines refers to our model mimicking a hard stop of the infalling material, dotted lines correspond to a more gradual turn-around of the ejecta (both described in section \ref{sec:ejecta}). 
}
\label{fig:trajt}
\end{figure}

\begin{figure}
\includegraphics[width=0.5\textwidth]{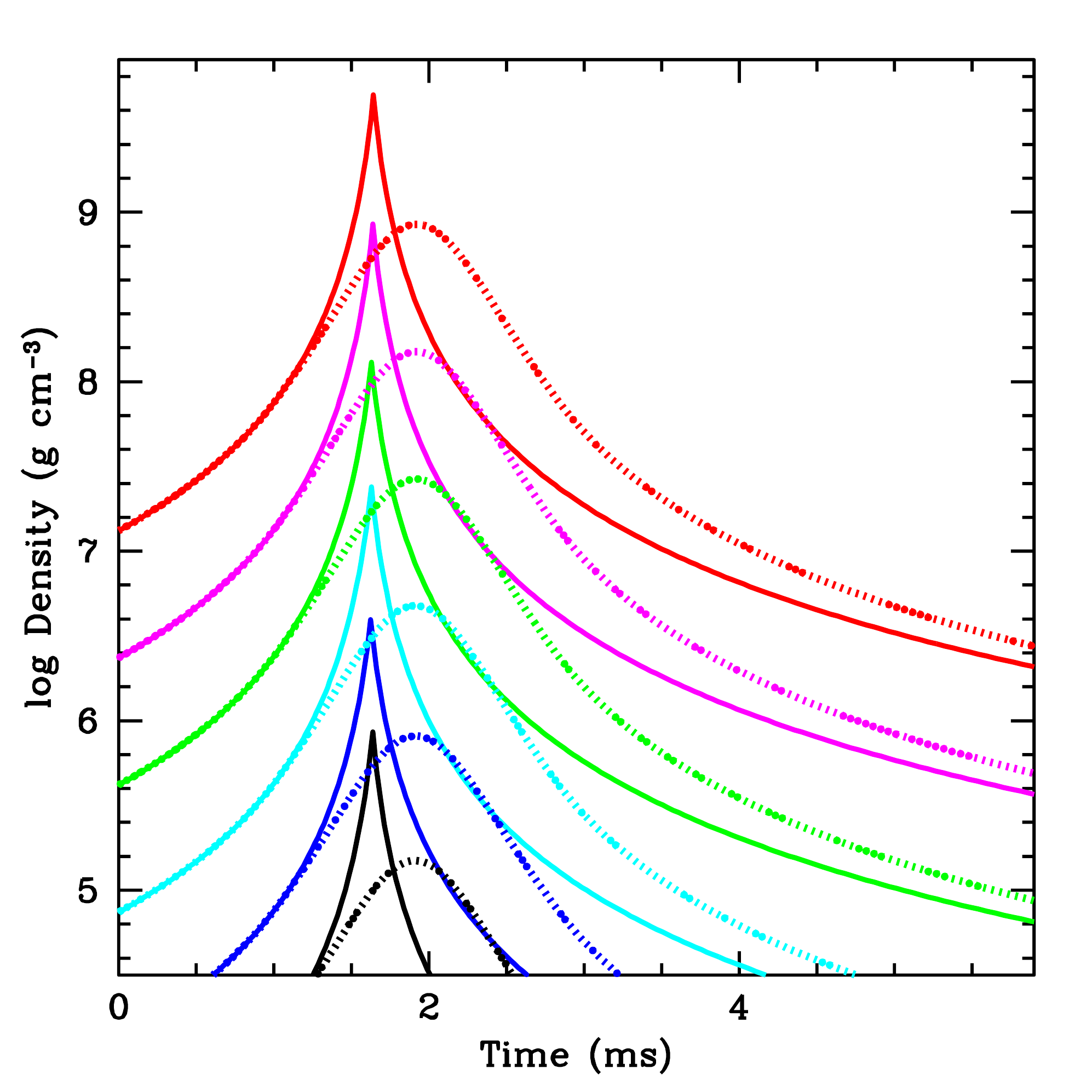} 
\caption{Density as a function of time for our trajectories at different accretion rates: 1$~\rm{{M}_{\odot}yr^{-1}}$ (black lines), 10$~\rm{{M}_{\odot}yr^{-1}}$ (blue lines), 10$^{2}~\rm{{M}_{\odot}yr^{-1}}$ (cyan lines), 10$^{3}~\rm{{M}_{\odot}yr^{-1}}$ (green lines), 10$^{4}~\rm{{M}_{\odot}yr^{-1}}$ (magenta lines) and 10$^{5}~\rm{{M}_{\odot}yr^{-1}}$ (red lines).  Solid lines refers to our model mimicking a hard stop of the infalling material, dotted lines correspond to a more gradual turn-around of the ejecta (both described in section \ref{sec:ejecta}). 
}
\label{fig:trajrho}
\end{figure}

The temperature evolution of these two paradigms is shown in Figure \ref{fig:trajt}.  {\rm The corresponding density evolution is shown in Figure \ref{fig:trajrho}. In the bounce trajectory, 
In our first model, sharp increase of the temperature evolution profile (Figure~\ref{fig:trajt}) mimics a hard stop for the infalling material prior to this expulsion.  In our the convective model, the acceleration begins sooner but is more gradual. }

Although we expect some gain in entropy during the heating phase, we assume constant entropy evolution for our models.  If we increase the entropy, we increase the temperature, and more heavy elements will be produced.  Likewise, if we make the evolution even more gradual, the material will remain in a region of high nuclear burning longer, also producing more heavy elements.  But our simulation-guided \citep{1996ApJ...460..801F,2006ApJ...646L.131F,2009ApJ...699..409F} toy models will provide a gauge of the importance of this ejecta in galactic chemical evolution.  

Typically $\sim$10-25\% of the accreted material is ejected along the angular momentum axis~\citep{2006ApJ...646L.131F}.  It is this material that is the focus of our nucleosynthetic studies.  In this paper, we focus on accretion rates between 1~$\rm{{M}_{\odot}yr^{-1}}$ and 10$^5~\rm{{M}_{\odot}yr^{-1}}$.  The rate rarely exceeds our highest accretion rate in common envelop situations and lower rates (occurring in initial phases of the CE) do not produce much nuclear burning.  The inner infall, nuclear burning and ejection phases are so rapid ($<1s$) that we can assume the accretion rate is constant over any cycle.

Table \ref{tab:summary one-zone} lists the trajectories investigated in this paper, and shown in Figure \ref{fig:trajt}. The trajectories labeled with ".C." are representative of the first model described above, while those labeled with ".D." are representative of the second delayed model.

\begin{table*}
\begin{tabular}{c|c|c|c|c|}
\toprule
Trajectory Name & Peak T$_{9}C$ & Trajectory Name & Peak T$_{9}D$  & Accretion Rate ($\rm{{M}_{\odot}yr^{-1}}$)\\
\midrule
mod.C.ar1d0 & 1.303 & mod.D.ar1d0 & 0.728 & 1 \\
mod.C.ar1d1 & 2.166 & mod.D.ar1d1 & 1.278 & 10\\
mod.C.ar1d2 & 3.981 & mod.D.ar1d2 & 2.306 & 10$^{2}$\\
mod.C.ar1d3 & 6.952 & mod.D.ar1d3 & 4.090 & 10$^{3}$\\
mod.C.ar1d4 & 10.00 & mod.D.ar1d4 & 7.286 & 10$^{4}$\\
mod.C.ar1d5 & 10.00 & mod.D.ar1d5 & 10.00 & 10$^{5}$\\
\bottomrule
\end{tabular}
\caption{List of trajectories investigated in this paper, along with peak temperatures in those trajectories. The first column includes those trajectories in Figure \ref{fig:trajt} which have undergone a sudden change in direction during the accretion, while those with a more gradual turnaround are listed in the third column. The accretion rates range from 1 to 10$^{5}~\rm{{M}_{\odot}yr^{-1}}$ as detailed in section \ref{sec:mdot}. Temperatures have been clipped in these models at 10 GK, as reaction rates tables above this threshold are not available.}
\label{tab:summary one-zone}
\end{table*}

\section{Nucleosynthesis calculations from Neutron Star Accretion}
\label{sec:yields}

In this section we present the nucleosynthesis yields of the trajectories described in section \ref{sec:ejecta} and listed in Table \ref{tab:summary one-zone}. 
The composition of the accreted material has a scaled solar isotopic distribution~\citep[][]{asplund:09}, with metallicity $Z_{m}=0.02$. 

The nuclear network includes 5234 isotopes and 74313 reactions from the different nuclear physics compilations and rates available \citep[see, e.g.,][]{pignatari:16}. The 3$-\alpha$ and $^{12}$C($\alpha$,$\gamma$)$^{16}$O by \citet{fynbo:05} and \citet{kunz:02} respectively, and the $^{14}$N(p,$\gamma$)$^{15}$O by \citet{imbriani:05}. The reaction rate $^{13}$C($\alpha$,n)$^{16}$O is taken from \citet{heil:08}, and the $^{22}$Ne($\alpha$,n)$^{25}$Mg and $^{22}$Ne($\alpha$,$\gamma$)$^{26}$Mg rates are from \citet{jaeger:01} and \citet{angulo:99}.
Experimental neutron capture reaction rates are taken when available from the KADoNIS compilation \citep[][]{dillmann:06}.  For neutron capture rates not included in KADoNIS, we refer to the JINA REACLIB database, V1.1 \citep[][]{cyburt:11}. The weak rates for light and intermediate-mass species are provided by \citet{oda:94} or \citet{fuller:85}, by \citet{langanke:00} for mass number 45 $<$ A $<$ 65 and for the weak interaction between protons and neutrons. Finally, for heavy species with A $>$ 65 we use \citet{goriely:99}.

Besides the $^{14}$N(p,$\gamma$)$^{15}$O mentioned above, proton capture rates have different sources like the NACRE compilation \citep[][]{angulo:99} and \cite{iliadis2001proton} for isotopes in the mass region between $^{20}$Ne and $^{40}$Ca. Proton captures on isotopes heavier than $^{40}$Ca are given by the JINA REACLIB database. The $^{13}$N(p,$\gamma$)$^{14}$O rate is taken from \cite{caughlan1988thermonuclear}, 
and proton captures on $^{27}$Al from \cite{champagne:92}.

For our simulations we have used material accreted with solar composition. 
This approximation does not affect conclusions in this paper. 
Indeed, at the lower accretion rates seed nuclei might have large effects on the distribution of products, but their impact on the final yields are marginal.
On the other hand, at the higher accretion rates investigated in this paper the material accreted enters in nuclear statistical equilibrium at extreme conditions (Table \ref{tab:summary one-zone}), so that the final yields are largely insensitive to the initial isotopic distribution (and to the electron fraction y$_{e}$). As we will see in the following sections, nucleosynthesis in these trajectories is dominating the total integrated ejecta for GCE. 

\subsection{Yields at Differing Accretion Rates} 
\label{sec: ppn set}

Production of isotopes at various accretion rates onto the surface of the compact object were investigated. Accretion rates varied from 1$~\rm{{M}_{\odot}yr^{-1}}$, where the neutron star first comes into contact with the companion, to the maximum accretion rate of 10$^{5}~\rm{{M}_{\odot}yr^{-1}}$, corresponding to the final stages of accretion in the system when the neutron star reaches the helium core.

Tables \ref{tab:tabulated_overproduction_mod_c} and  \ref{tab:tabulated_overproduction_mod_d} summarize the results of this section and give a list of the isotopes with the highest overproduction, the increase in the abundance of an isotope relative to its initial abundance in the accreted material, at the accretion rates 10$^{4}~\rm{{M}_{\odot}yr^{-1}}$ and 10$^{5}~\rm{{M}_{\odot}yr^{-1}}$ for the trajectories investigated. Trajectories at lower accretion rates do not show the same extent of overproduction, and therefore contribute minimally to our GCE results. Table \ref{tab:most_overproduced_tab} shows the top 5 isotopes overproduced in each of the trajectories investigated. Those isotopes most produced in these trajectories are also given in the relevant sections.
In Tables \ref{tab:centraj_abunds} and \ref{tab:delayed_abundances_table} the isotopic yields of stable isotopes (with contributions from decayed unstable isotopes) are provided for the trajectories listed in Table \ref{tab:summary one-zone}. Complete radiogenic contribution is considered for these abundances. Undecayed abundances are also provided in separate tables available on-line at \href{http://apps.canfar.net/storage/list/nugrid/nb-users/Common_Envelope_Data_Keegans_2018}{CANFAR}. 
The corresponding overproduction factors are given in Table \ref{tab:overprod_factors_centraj} and table \ref{tab:overprod_factors_delay}.

Complete isotopic production factors and elemental abundances are shown in Figures \ref{fig:overabunds_multii} and \ref{fig:mass_frac_multii}, respectively, where results from the two sets of trajectories at accretion rates can be compared (see discussion here below). Figure \ref{fig:flux_multi_figure} shows the integrated fluxes for each of the Mod.C accretion rates.
Figures \ref{fig:multiplot_element_overabnd_with_legend} and \ref{fig:multiplot_element_overabnd_edlayed_with_legend} show the final isotopic production factors, zoomed in the mass region 40 $\lesssim$ A  $\lesssim$ 100, and the electron fraction Ye obtained. Our results show that the largest contributions to GCE must come from those trajectories at higher accretion rates. In these conditions, hydrogen is fully burned allowing high abundance overproductions at the iron group and beyond. Lower accretion rates contribute marginally to the enrichment of ejected material, both because the material is incompletely burnt and the amount of material ejected is much smaller. Proton rich material is highly overproduced in trajectory mod.C.ar1d4, and this is the source of most of the proton rich material ejected from the system. This trajectory reaches high enough temperatures that the accreted fuel burns efficiently, with a low enough peak density such that electron captures do not become dominant causing a shift in the $Y_\mathrm{e}$ and peak production to more neutron rich isotopes.


\begin{table}
\begin{tabular}{p{2cm}|p{2.6cm}|p{2.6cm}|}
\toprule
Overproduction $x_{fin}/x_{ini}$ & 10$^{4}~\rm{{M}_{\odot}yr^{-1}}$ & 10$^{5}~\rm{{M}_{\odot}yr^{-1}}$\\
\midrule
10$^{4} < x < 10^{5} $&
$^{76,77}$Se,
$^{60,61}$Ni,
$^{72,73}$Ge,
$^{86,87}$Sr,
$^{75}$As,
$^{82}$Kr,
$^{69,71}$Ga,
$^{63,65}$Cu,
$^{96}$Ru,
$^{89}$Y,
$^{66,67,68}$Zn,
$^{94}$Mo,
& $^{84}$Kr,
$^{67,68,70}$Zn,
$^{63,65}$Cu,
$^{89}$Y,
$^{58}$Fe,
$^{86}$Kr, \\
10$^{5} < x < 10^{6} $ & 
$^{70}$Ge,
$^{64}$Zn,
$^{80}$Kr,
$^{93}$Nb,&
$^{88}$Sr,
$^{87}$Rb,
$^{62,64}$Ni\\
10$^{6} < x < 10^{7} $ & $^{74}$Se,~$^{78}$Kr, $^{84}$Sr,~$^{90,91}$Zr& - \\
10$^{7} < x          $ & $^{92}$Mo & - \\
\bottomrule
\end{tabular}
\caption{
Table showing those isotopes overproduced in the given ranges for each of the accretion rates shown in the mod.C trajectories
}
\label{tab:tabulated_overproduction_mod_c}
\end{table}


\begin{table}
\begin{tabular}{p{2cm}|p{2.6cm}|p{2.6cm}|}
\toprule
Overproduction $x_{fin}/x_{ini}$ & 10$^{4}~\rm{{M}_{\odot}yr^{-1}}$ & 10$^{5}~\rm{{M}_{\odot}yr^{-1}}$\\
\midrule
10$^{4} < x < 10^{5} $ &
$^{58,61}$Ni &
$^{76}$Ge,
$^{50}$Ti,
$^{61,62}$Ni,
$^{63,65}$Cu,
$^{82}$Se,
$^{66,67,68,70}$Zn,
$^{58}$Fe, \\
10$^{5} < x < 10^{6} $ &
- &
$^{88}$Sr,
$^{87}$Rb,
$^{64}$Ni,
$^{86}$Kr,
$^{54}$Cr \\
\bottomrule
\end{tabular}
\caption{
Table showing those isotopes overproduced in the given ranges for each of the accretion rates shown in the mod.D trajectories
}
\label{tab:tabulated_overproduction_mod_d}
\end{table}

\begin{table}
\begin{tabular}{p{3.0cm}|p{4.5cm}|}
\toprule
Model  & Most overproduced isotope \\
\midrule
mod.C.ar1d0  & $^{42}$Ca, $^{21}$Ne, $^{74}$Se, $^{18}$O, $^{ 7}$Li \\
mod.C.ar1d1  & $^{32}$Ba, $^{80}$Ta, $^{98}$Ru, $^{31}$P, $^{38}$La \\
mod.C.ar1d2  & $^{58}$Ni, $^{98}$Ru, $^{62}$Ni, $^{51}$V, $^{61}$Ni \\
mod.C.ar1d3  & $^{64}$Zn, $^{51}$V, $^{58}$Ni, $^{62}$Ni, $^{61}$Ni \\
mod.C.ar1d4  & $^{91}$Zr, $^{90}$Zr, $^{78}$Kr, $^{74}$Se, $^{92}$Mo\\
mod.C.ar1d5  & $^{86}$Kr, $^{62}$Ni, $^{64}$Ni, $^{87}$Rb, $^{88}$Sr, \\
mod.D.ar1d0  & $^{31}$P, $^{33}$S, $^{15}$N, $^{18}$O, $^{ 7}$Li \\
mod.D.ar1d1  & $^{23}$Na, $^{33}$S, $^{42}$Ca, $^{21}$Ne, $^{ 7}$Li \\
mod.D.ar1d2  & $^{36}$Ce, $^{41}$K, $^{44}$Ca, $^{43}$Ca, $^{42}$Ca \\
mod.D.ar1d3  & $^{64}$Zn, $^{51}$V, $^{58}$Ni, $^{61}$Ni, $^{62}$Ni \\
mod.D.ar1d4  & $^{64}$Zn, $^{60}$Ni, $^{62}$Ni, $^{58}$Ni, $^{61}$Ni \\
mod.D.ar1d5  & $^{54}$Cr, $^{88}$Sr, $^{86}$Kr, $^{87}$Rb, $^{64}$Ni \\
\end{tabular}
\caption{5 most overproduced isotopes for each of the accretion rates investigated.}
\label{tab:most_overproduced_tab}
\end{table}

\subsubsection{Accretion rate 1 $\rm{{M}_{\odot}yr^{-1}}$: mod.C.ar1d0 and mod.D.ar1d0} 
\label{1_M_dot_sec}

The dominant reactions are proton captures from the H rich accreted material, followed by $\beta^{+}$ decays bringing the material back towards the valley of stability in the mod.C.ar1d0 case. Some (p,$\alpha$) reactions are evident in this region, hindering the flow of
material to heavier masses.

Significant increases in the abundance of $^{7}$Li can be seen for both mod.C.ar1d0 and mod.D.ar1d0 (top left panel of Figure \ref{fig:overabunds_multii}) with both trajectories showing overproduction factors of $\sim$10$^{6}$. This occurs due to efficient $^{3}$He ($\alpha,\gamma$)$^{7}$Be, followed by {$\beta^{+}$} decays. $^{15}$N and $^{18}$O are also increased significantly, by factors of about 150 and 250 for mod.C.ar1d0, and 280 and 510 for the delayed trajectory. Enhancements of light and intermediate-mass elements
are comparable for the two trajectories, until the mass region A $\sim$ 50. Beyond this mass, the delayed trajectory does not show efficient production for the heavier elements, while in mod.C.ar1d0 final abundances are enhanced up to A $\sim$ 80. This is due to the higher peak temperature reached compared to the mod.D.ar1d0 model (1.303 GK compared to 0.728 GK), causing a build up of heavier nuclei mostly via proton capture reactions.

Panel a of Figure \ref{fig:mass_frac_multii} shows that neither of the trajectories at this accretion rate allows any burning of material above Z=40. The conditions at this accretion rate are not extreme enough to allow proton captures on these heavier nuclei, or to cause photo-disintegration on the heavier elements present in the accreted material.

$^{74}$Se, $^{78}$Kr and $^{84}$Sr (the lightest of the p-nuclei) all have overproduction factors greater than 10$^{2}$ for mod.C.ar1d0 at this accretion rate, where as the delayed trajectory shows no production of these isotopes. Inspection of flux charts for mod.C.ar1d0 shows a proton capture path with $\beta^{+}$ decays in the proton-rich side of the valley of stability, like for the rp-process \citep[][]{schatz:01}. 
This becomes most evident above the N=20 neutron magic number (see panel (a) of Figure \ref{fig:flux_multi_figure}). As can be seen from the first columns of Tables \ref{centraj_abundances_table} and \ref{tab:overprod_factors_delay}, the H fuel accreted from the companion star remains largely unburnt under these conditions - only 0.2\% of the initial abundance of H is burnt for both of these trajectories.


\subsubsection{Accretion rate 10$~\rm{{M}_{\odot}yr^{-1}}$: mod.C.ar1d1 and mod.D.ar1d1} 
\label{10_M_dot_sec}

Nucleosynthesis in mod.C.ar1d1 results in overproduction of a large number of isotopes up to A$\sim$157, due to the higher peak temperatures experienced in this trajectory as compared with the mod.C.ar1d0 case, allowing for more efficient proton captures in the accreted material, whilst remaining below the threshold for activation of photodisintegration of heavier material seen in higher accretion rate models. The highest mass isotope to be overproduced under these conditions is $^{180}$Ta (with a production factor of 5.4$\times $10$^{2}$), and the isotopes with the greatest overproduction are $^{138}$La and $^{31}$P (3.3$\times $10$^{3}$ and 1.0$\times $10$^{3}$ times initial abundance respectively). 14 other isotopes - $^{33}$S, $^{35, 37}$Cl, $^{42}$Ca, $^{62}$Ni, $^{84}$Sr, $^{96, 98}$Ru, $^{102}$Pd, $^{120}$Te, $^{126}$Xe, $^{130, 132}$Ba and $^{136, 138}$Ce all have production factors of between 10$^{2}$ and 10$^{3}$.

In mod.D.ar1d1 abundances are greatest at lower masses. $^{7}$Li, $^{21}$Ne, $^{23}$Na, $^{33}$S, $^{42}$Ca and $^{84}$Sr have overproduction factors of between 10$^{2}$ and 10$^{3}$. The much reduced range of highly overproduced isotopes is again due to the lower temperatures experienced by the delayed trajectory on in-fall of the material. The peak temperature of mod.D.ar1d1 is similar to the peak temperature of mod.C.ar1d0 trajectory discussed before (see Table \ref{tab:summary one-zone}). Differences in the two abundance distributions are due to the temperature histories of the two trajectories, with the delayed trajectory exposed for longer to more  extreme conditions.

\begin{figure*}
\includegraphics[width=1\textwidth]{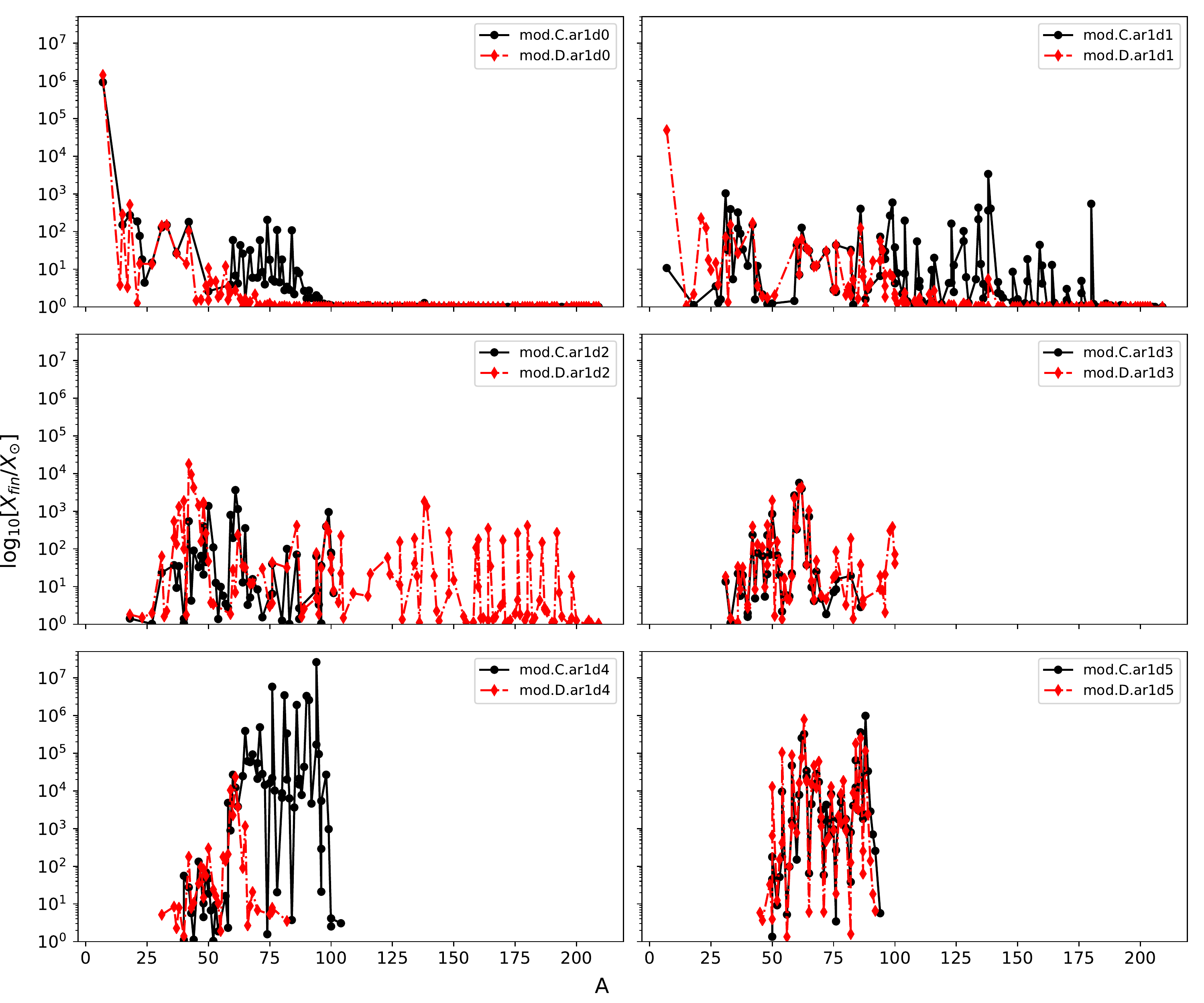}
\caption{ 
Overproduction factors for trajectories listed in Table \ref{tab:summary one-zone} are shown. Abundances for delayed trajectories ("mod.D.") are shown in red, and in black for "mod.C." trajectories. Relative accretion rates are indicated in each panel. 
}
\label{fig:overabunds_multii}
\end{figure*}

\begin{figure*}
\includegraphics[width=1\textwidth]{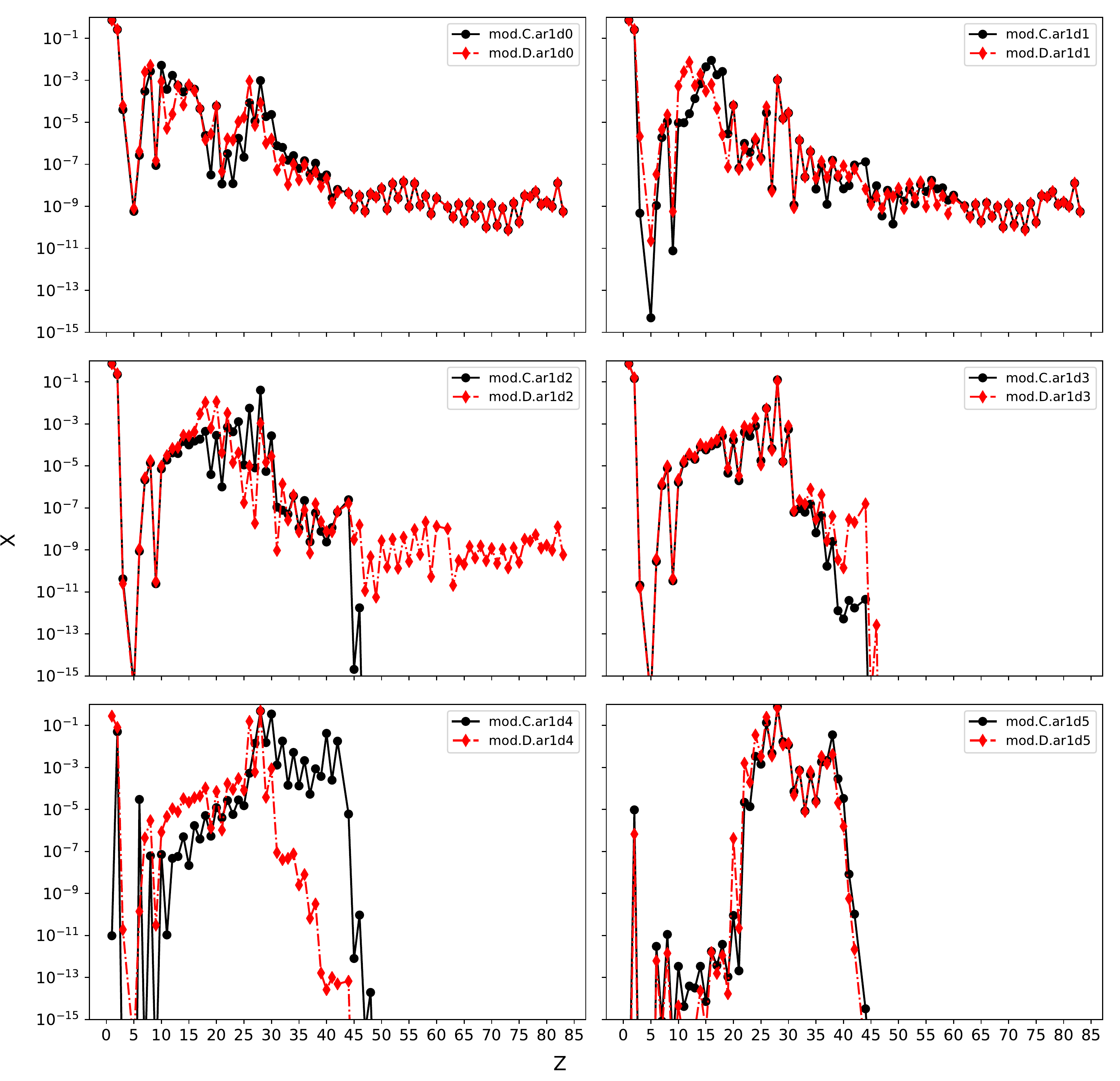}
\caption{ 
Element abundance distributions in mass fraction with respect to atomic number are shown for the same cases in figure \ref{fig:overabunds_multii}. Contribution from radioactive decay is considered. Abundances for delayed trajectories ("mod.D.") are shown in red, and in black for "mod.C." trajectories. Most of the element abundances in the first two panels remains unchanged from their initial distributions, while significant changes can be observed at higher accretion rates.
}
\label{fig:mass_frac_multii}
\end{figure*}

Production factors and abundances are shown in the top right panels of Figures \ref{fig:overabunds_multii} and \ref{fig:mass_frac_multii}, respectively. The mass fraction of the majority of stable isotopes above Z=40 remains unchanged, as was the case for the previous section. Peak temperatures from our simulations for these trajectories do not increase beyond $\approx$2 GK (Figure \ref{fig:trajt}) and burning occurs on a timescale of order milliseconds. Neither $\alpha$ nor proton captures have high enough probability at these temperatures to trigger complete burning of the accreted material.

\begin{figure*}
	\centering
    \subfigure[]{\label{fig:centraj_1_flux_image}}\includegraphics[clip, trim = 9cm 0.2cm 2cm 1cm, width=0.45\textwidth]{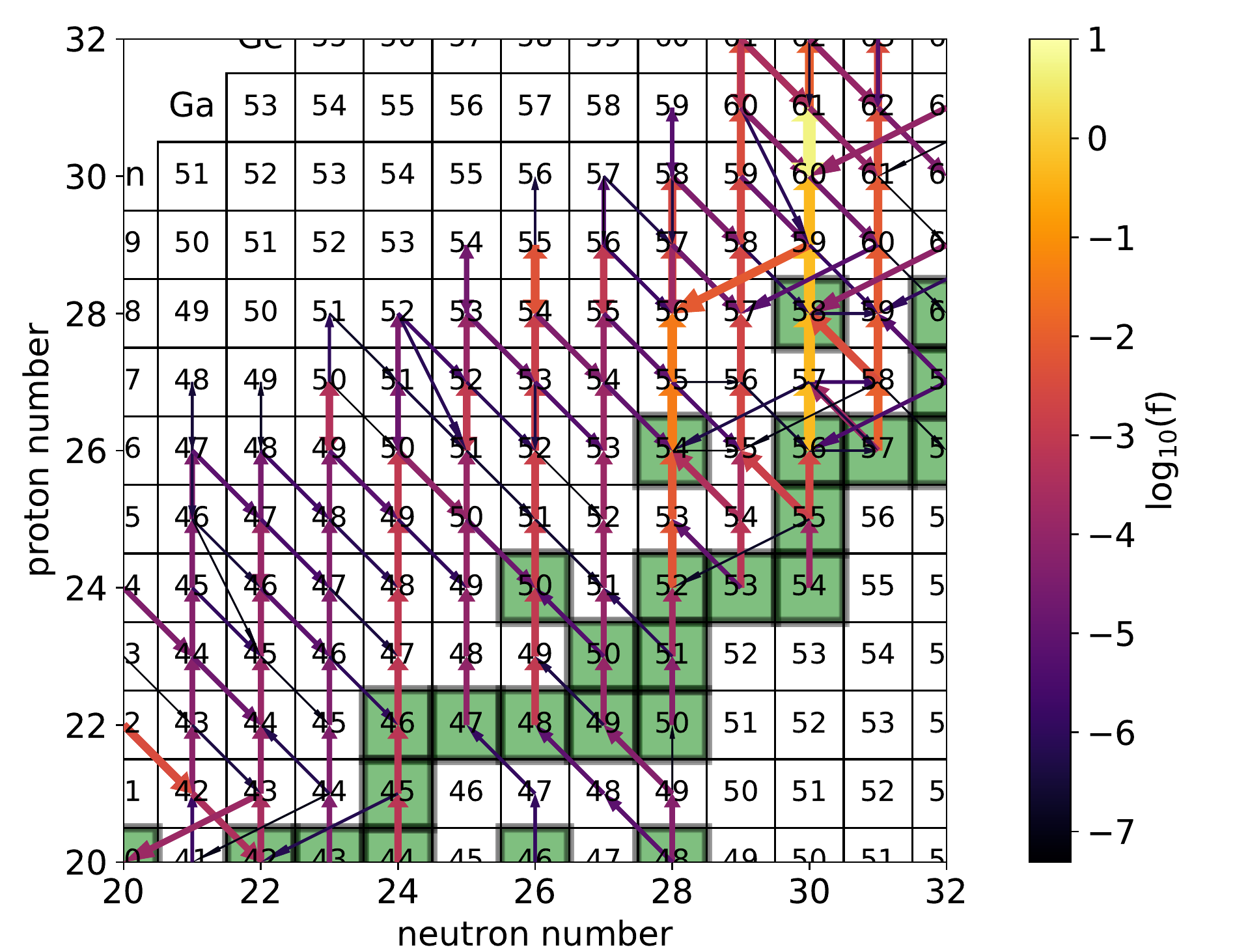}
    \subfigure[]{\label{fig:b}}\includegraphics[clip, trim = 9cm 0.2cm 2cm 1cm, width=0.45\textwidth]{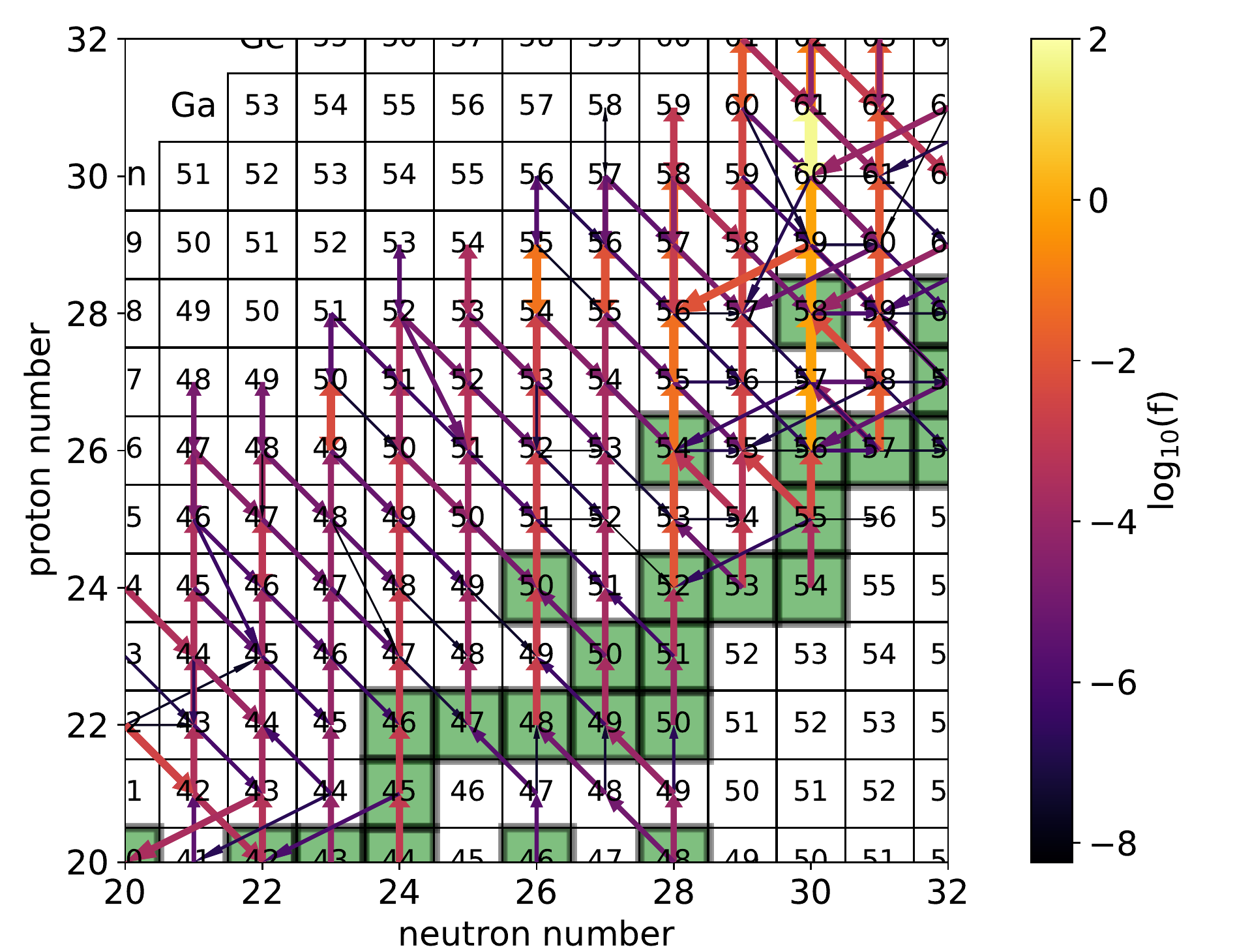}
    \subfigure[]{\label{fig:b}}\includegraphics[clip, trim = 9cm 0.2cm 2cm 1cm, width=0.45\textwidth]{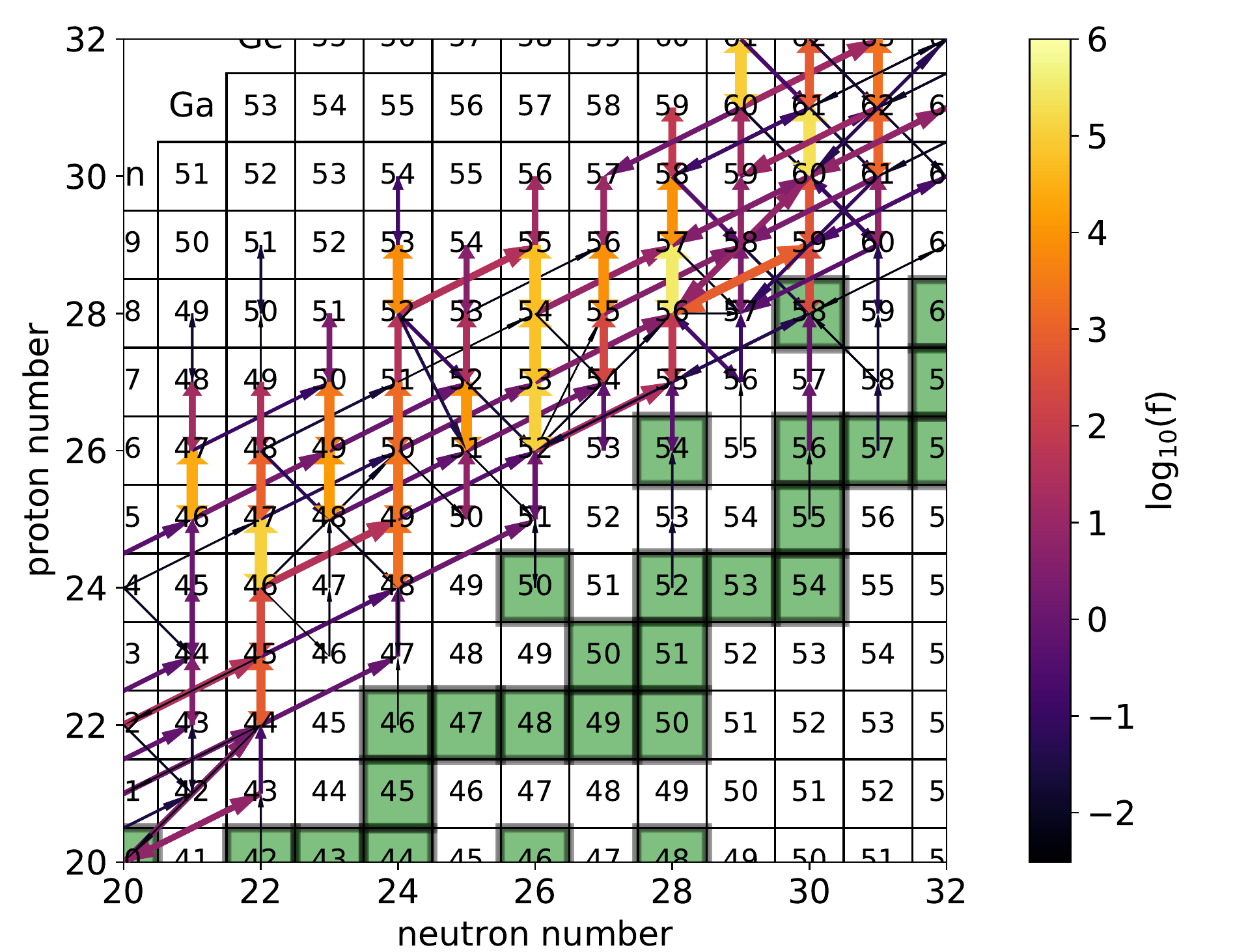}
    \subfigure[]{\label{fig:b}}\includegraphics[clip, trim = 9cm 0.2cm 2cm 1cm, width=0.45\textwidth]{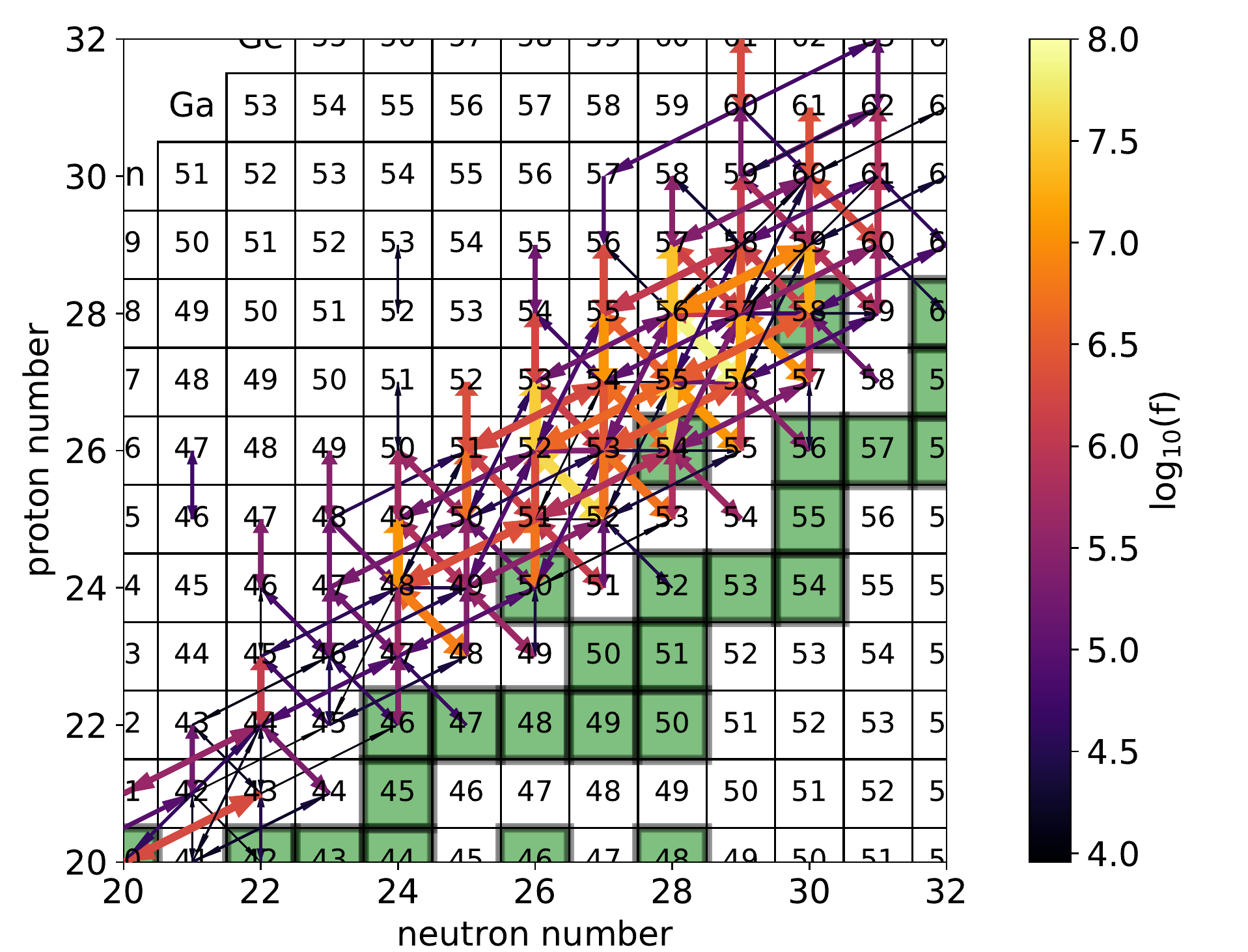}
    \subfigure[]{\label{fig:b}}\includegraphics[clip, trim = 9cm 0.2cm 2cm 1cm, width=0.45\textwidth]{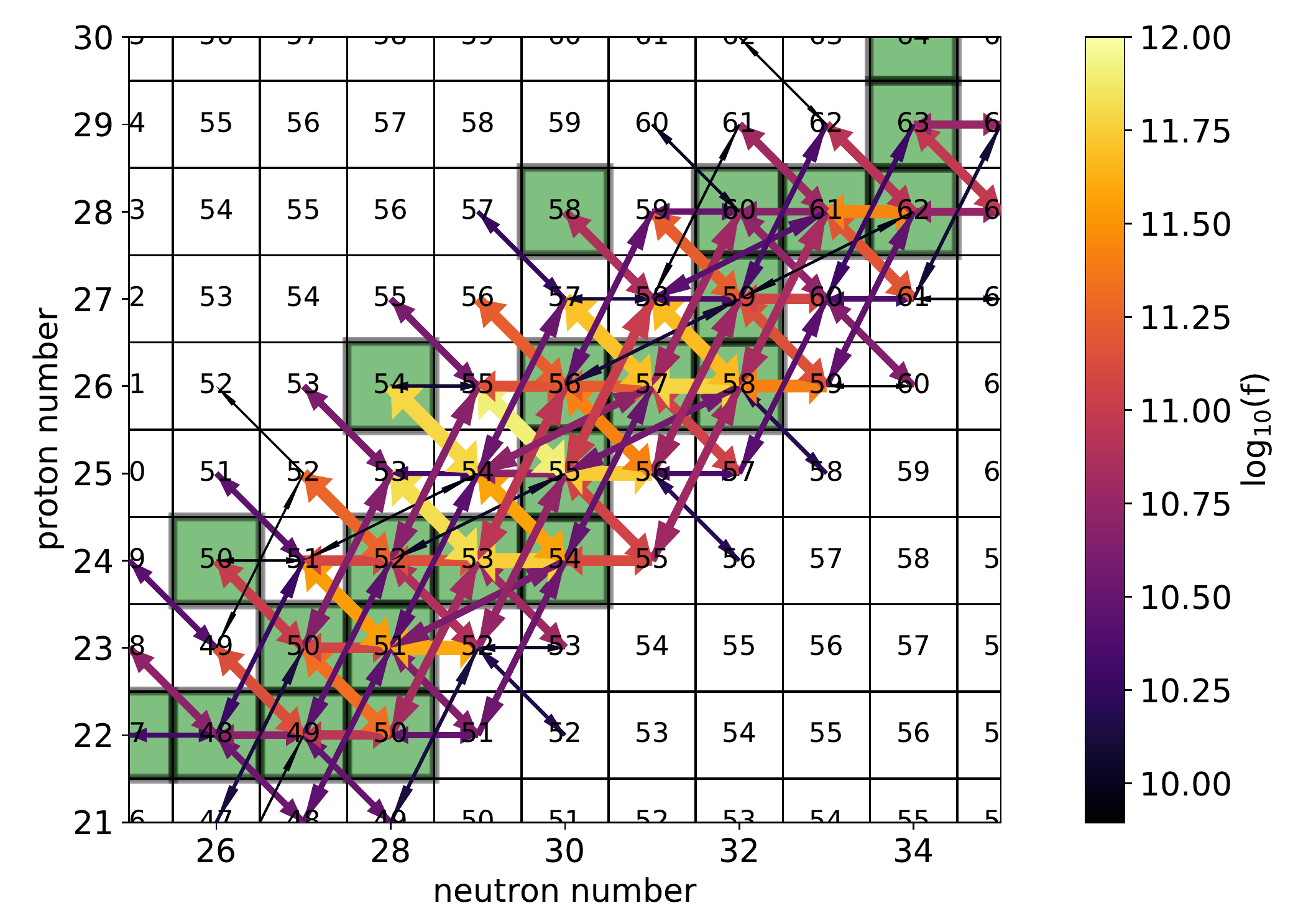}
    \subfigure[]{\label{fig:b}}\includegraphics[clip, trim = 9cm 0.2cm 2cm 1cm, width=0.45\textwidth]{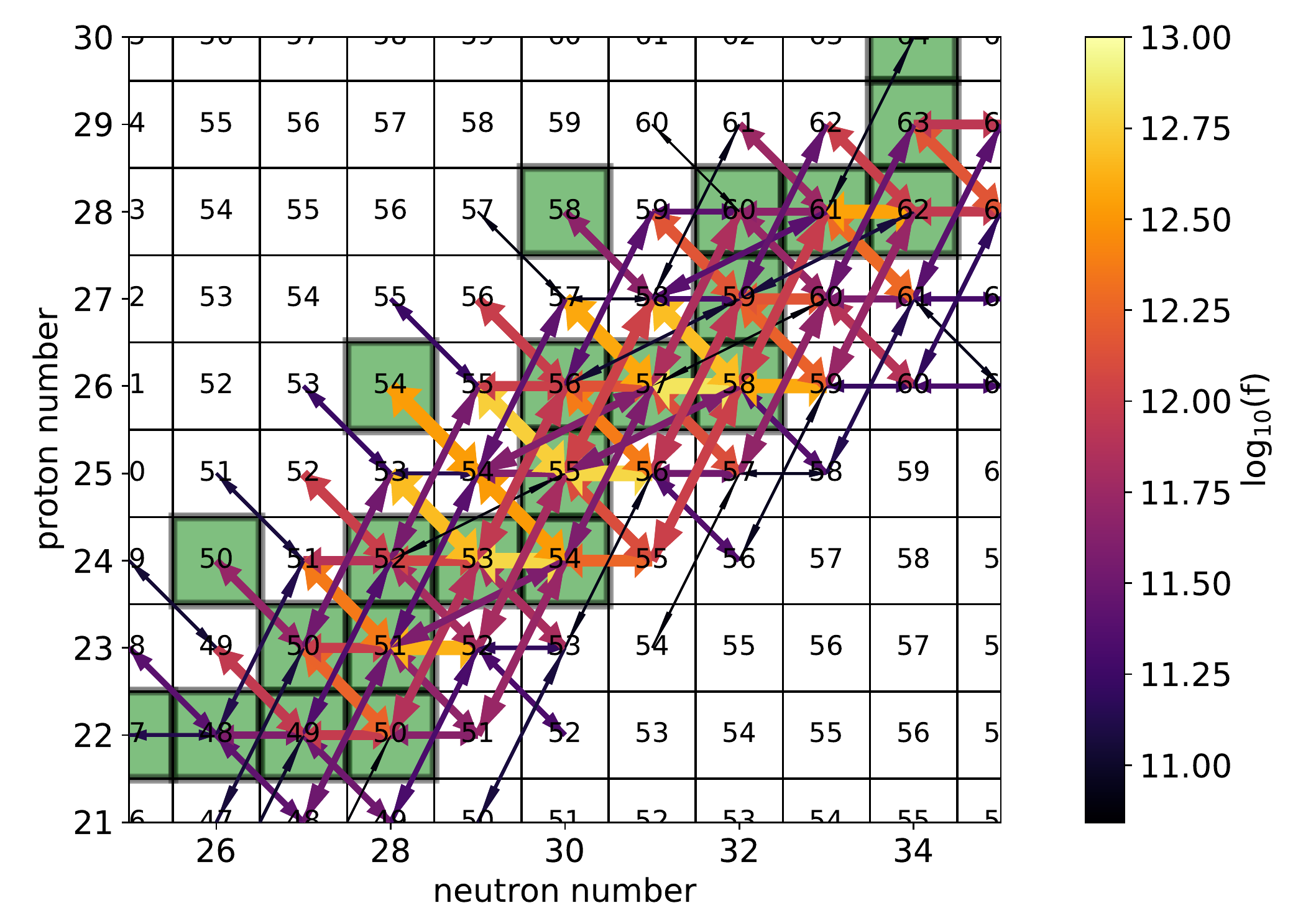}
\caption{Figure showing the distribution of fluxes ($\delta X/ \delta t)$) for each reaction ( units s$^{-1}$) for the mod.C trajectories, normalised to the maximum flux in each of those trajectories}. Top left to bottom, these correspond to the different accretion rates investigated in this paper for the mod.C case, lowest accretion rate to highest.
\label{fig:flux_multi_figure}
\end{figure*}

Flux nucleosynthesis plots for this accretion rate are similar to those for the 1$~\rm{{M}_{\odot}yr^{-1}}$ case, and are shown in panel (b) of Figure \ref{fig:flux_multi_figure}.

\subsubsection{Accretion rate 10$^{2}~\rm{{M}_{\odot}yr^{-1}}$: mod.C.ar1d2 and mod.D.ar1d2} 
\label{100_M_dot_sec}

Abundance production for mod.C.ar1d2 is clustered at mass between A $\sim$ 25 and 100. $^{51}$V and $^{61,62}$Ni are overproduced by factors larger than 10$^{3}$. Eight other isotopes have production factors greater than 10$^{2}$: $^{42}$Ca, $^{48}$Ti, $^{52}$Cr, $^{58, 60}$Ni, $^{64}$Zn and $^{96, 98}$Ru. 

The flux nucleosynthesis plot (panel (c) of Figure \ref{fig:flux_multi_figure}) for this trajectory shows a relevant difference compared to lower accretion rates. Burning proceeds further from the valley of stability, due to the more efficient proton captures at high temperatures. The $\alpha$-captures on intermediate-mass isotopes are activated. Heavier isotopes are destroyed by ($\gamma$,n) photodisintegration reactions, and a mild flux of (n,$\gamma$) reactions is also activated. As for lower accretion rates, a large proportion of the infalling material is not burnt (Table \ref{tab:centraj_abunds}) with only $\approx$0.3\% of the hydrogen fuel being consumed.  The peak temperatures in this trajectory is 3.981 GK.


The delayed trajectory mod.D.ar1d2 shows an abundance distribution extremely different from mod.C.ar1d2. 
Production extends to higher mass isotopes due to the longer time that the material spends at high temperature, and without yet relevant activation of photo-disintegration reactions. There are 22 isotopes enhanced by a factor between 10$^{2}$ and 10$^{3}$,  and among them 8 have masses above A = 150. $^{38}$Ar, $^{41}$K, $^{43,44}$Ca, $^{45}$Sc, $^{47,48}$Ti and $^{136,138}$Ce show overabundances between 10$^{3}$ and 10$^{4}$, and $^{42}$Ca greater than 10$^{4}$. Temperatures are not high enough to initiate the photodisintegration reactions observed in mod.C.ar1d2.

It can be seen from the middle left panel of Figure \ref{fig:mass_frac_multii} that there is a significant change in the mass fraction of elements with 50 < Z < 65, although the very heaviest elements (with Z > 65) which are investigated in this work remain largely unchanged in their abundances - The distributions of both of the C and D model are identical in the upper panels of Figure \ref{fig:mass_frac_multii}, implying that the distribution cannot be changed from the initial abundances. This is verified by comparison with Figure \ref{fig:overabunds_multii}, where the production factors for all isotopes above A = 150 and most above A=100 are negligible. Some elements with 50 > Z > 65 have abundances increased by up to 3 orders of magnitude. However, because of their small intrinsic abundances and the small contribution to the final integrated ejecta, this effect is not important to GCE models.


\subsubsection{Accretion rate 10$^{3}~\rm{{M}_{\odot}yr^{-1}}$: mod.C.ar1d3 and mod.D.ar1d3} 
\label{1000_M_dot_sec}

Material at this accretion rate reaches a peak temperature of 6.952 GK in the mod.C.ar1d3 trajectory and 4.090 GK in the mod.D.ar1d3 case, although for both trajectories the accreted material remains largely unburnt. The trajectory mod.C.ar1d3 reaches nuclear statistical equilibrium, and the destruction of heavier elements by photodisintegration as already seen for mod.C.ar1d2 now occurs in both trajectories. Isotopes are produced most efficiently in the mass region 50 $\lesssim$ A  $\lesssim$ 100, including species in the iron-group region and up to Ru.
For both trajectories, a significant increase of greater than 10$^{3}$ in the abundances of $^{58,61,62}$Ni is seen. 
In mod.C.ar1d3 overabundances greater than 10$^{2}$ are obtained for 5 other isotopes: $^{48}$Ti, $^{42}$Ca, $^{60}$Ni, $^{64}$Zn and $^{51}$Va, and 9 in the delayed trajectory mod.D.ar1d3: $^{45}$Sc, $^{42,44}$Ca, $^{48,49}$Ti, $^{52}$Cr, $^{60}$Ni, $^{80}$Kr and $^{96}$Ru. 
Elemental distributions of the two trajectories for this accretion rate are similar (middle left panel of Figure~\ref{fig:mass_frac_multii}) up to Z = 30, however there are significant differences in the abundances of Mo, Nb and Ru, none of which are overproduced in mod.C.ar1d3, and for some isotopes of Kr and Sr where a difference in overproduction factors of over an order of magnitude can be observed.

In panel (d) of figure \ref{fig:flux_multi_figure}, the integrated reaction flows are shown for mod.C.ar1d3. In these conditions, $\alpha$ captures and $\beta$ decays are now the dominant reaction pathways. Proton captures are still active, but they do not push material from the line of stability as efficiently as in panel (c) of Figure \ref{fig:flux_multi_figure}. Approximately 50\% of the initial abundance of He is burnt in this trajectory.  

 

\subsubsection{Accretion rate 10$^{4}~\rm{{M}_{\odot}yr^{-1}}$: mod.C.ar1d4 and mod.D.ar1d4} 
\label{10000_M_dot_sec}

Among the trajectories described in this section, mod.C.ar1d4 is the first case to undergo complete burning of hydrogen, with this trajectory entering nuclear statistical equilibrium (NSE). Significant overproduction for a large number of isotopes is observed: $^{74}$Se, $^{78}$Kr, $^{84}$Sr and $^{90,91}$Zr have final abundances over 6 orders of magnitude higher than initial, $^{92}$Mo is overproduced by a factor of 2.6$\times10^{7}$. Excepting the Zr isotopes, all of these nuclei are classically defined as products of the p-process \citep[][]{arnould:03,rauscher:13,pignatari:16}. While these are proton rich isotopes, for mod.C.ar1d4 we obtain a final neutron rich distribution (Ye  = 0.46, table \ref{tab:centraj_abunds}). This is mostly due to the neutron-rich abundance signature in the Ni region, with 58\% of mass fraction as $^{62}$Ni and only a minor contribution to $^{58}$Ni. The zoomed isotopic distribution is shown in Figure \ref{fig:multiplot_element_overabnd_with_legend}, bottom left panel. The abundance pattern of heavy isotopes is similar to neutrino-driven winds ejecta with proton-rich composition \citep[e.g.,][]{froehlich:06,roberts:10,arcones:11}.

Partial burning of the hydrogen is observed for the delayed trajectory mod.D.ar1d4. Production
for heavy isotopes is marginal compared to mod.C.ar1d4 (Figure \ref{fig:overabunds_multii}),
while intermediate-mass elements are made more efficiently (Figure
\ref{fig:mass_frac_multii}). Isotopes $^{58,61}$Ni both have overproduction factors greater
than 10$^{4}$. As also shown in the bottom left panel of Figure
\ref{fig:multiplot_element_overabnd_edlayed_with_legend}, p-process isotopes are not made in
mod.D.ar1d4, with no significant production above A $\sim$ 80. As we have seen for trajectories
in the previous section, material above this mass has been destroyed by photodisintegration.
Panel (e) of Figure \ref{fig:flux_multi_figure} shows the flux plot for mod.C.ar1d4. 
The material is in NSE,
with the large amount of hydrogen allowing proton captures to occur and extending the
abundance distribution to the proton rich side of the line of stability. The lower
densities reached in this model as compared with the mod.C.ar1d5 trajectory ensure that electron
capture reactions are not favorable enough to reduce the $Y_\mathrm{e}$. 

The trajectory is in NSE with large fluxes in the iron group region, lower average and lower peak densities in this model lead to a more proton rich distribution of products than the mod.C.ar1d5 case (see panel f of Figure \ref{fig:flux_multi_figure} for comparison). The flux plot is dominated by the reactions in NSE equilibrium around the iron group region during the NSE phase. In these simulations, the temperature and density freezout is extremely fast and do not allow to significantly modify the integrated fluxes after the trajectories exit from NSE. 



\subsubsection{Accretion rate 10$^{5}~\rm{{M}_{\odot}yr^{-1}}$: mod.C.ar1d5 and mod.D.ar1d5} 
\label{100000_M_dot_sec}

More neutron rich material is produced under these conditions with isotopic distributions skewed towards more neutron rich isotopes. This is due to the higher peak and average density conditions experienced in this model as compared with the mod.C.ar1d4 trajectory. The most overproduced isotope in trajectory mod.C.ar1d5 is $^{88}$Sr, with an overproduction factor of 9.8$\times$10$^{5}$ along with $^{87}$Rb and $^{62,64}$Ni, all with overproduction factors of greater than 10$^{5}$. A similar distribution is observed in the mod.D.ar1d5 model, however $^{64}$Ni is the most overproduced isotope in this case, with $^{87}$Rb, $^{86}$Kr, $^{88}$Sr and $^{54}$Cr being the other isotopes with a production factor greater than 10$^{5}$. A large number of isotopes in both the mod.C.ar1d5 and mod.D.ar1d5 models are overproduced by factors of $\sim$ 10$^{4}$ - 9 for the mod.C.ar1d5 case and 12 for the mod.D.ar1d5, all of which are clustered around the iron group region 
Temperatures in mod.C.ar1d5 have been clipped at 10 GK as reaction rate tables 
are not available beyond this value.

\begin{figure*}
\includegraphics[width=1\textwidth]{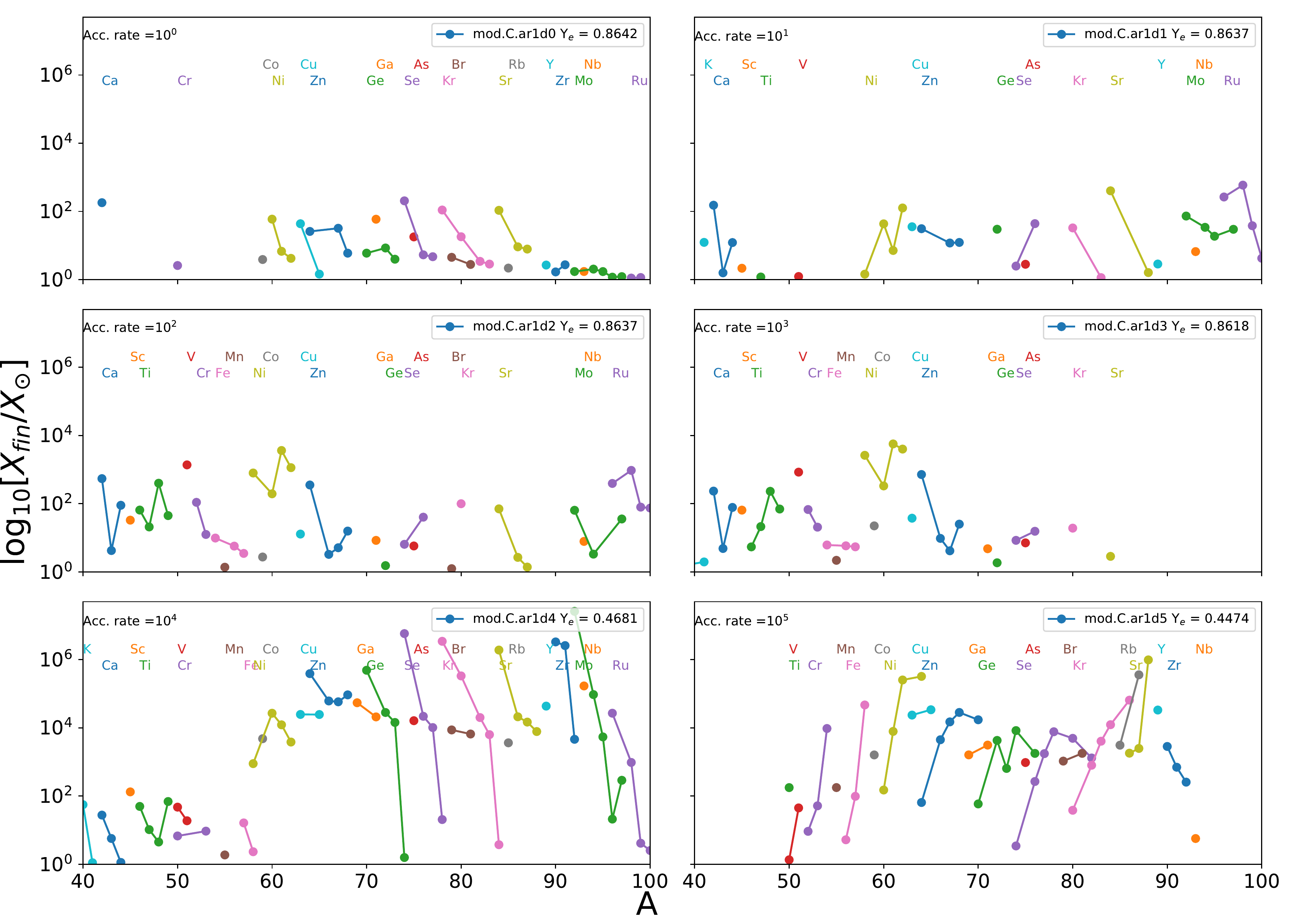}
\caption{The isotopic distribution between A = 40 and 100 is shown for trajectories "mod.C." at different accretion rates. 
}
\label{fig:multiplot_element_overabnd_with_legend}
\end{figure*}

\begin{figure*}
\includegraphics[width=1\textwidth]{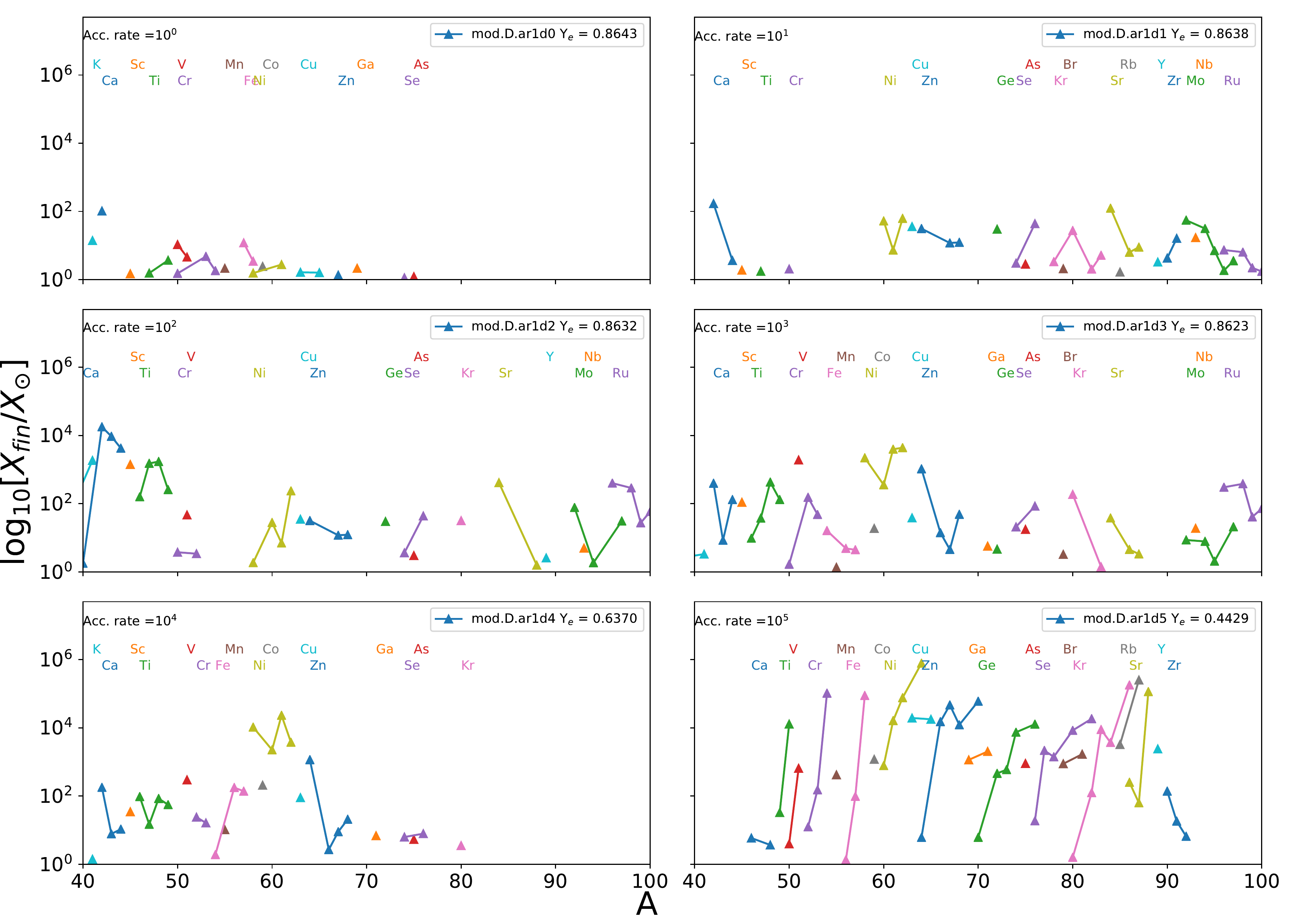}
\caption{As for Figure \ref{fig:multiplot_element_overabnd_with_legend} but for delayed trajectories ("mod.D."). 
}
\label{fig:multiplot_element_overabnd_edlayed_with_legend}
\end{figure*}

The overall distribution and range of production is similar between mod.C.ar1d5 and mod.D.ar1d5, as can be seen in the bottom right panel of Figure \ref{fig:mass_frac_multii} and Figures \ref{fig:multiplot_element_overabnd_with_legend} and \ref{fig:multiplot_element_overabnd_edlayed_with_legend}.

As discussed by \cite{2006ApJ...646L.131F}, fall-back trajectories can produce r-process abundances with mild neutron-rich conditions. In the scenario discussed here, within a realistic range of accretion rates, mod.C.ar1d5 and mod.D.ar1d5 both show neutron-rich $Y_\mathrm{e}$ (0.447 and 0.443, respectively) and an abundance signature similar to the weak r-process \citep[e.g.,][]{seeger:65,kratz:93,arcones:11,wanajo:13}. Collapsars, as investigated by \cite{siegel2018neutron}, with accretion rates from ~0.3 to 30 times those investigated in these ar1d5 trajcectories have also been shown to have an increase in neutron rich material.

Panel (f) of Figure \ref{fig:flux_multi_figure} shows the integrated flux plot for the mod.C.ar1d5 trajectory. It can be seen that the reaction pathways for this trajectory are through more neutron rich isotopes than in the mod.C.ar1d4 case. This is due to the higher density over the period of infall and ejection, making electron capture decays more efficient. In contrast to the mod.C.ar1d4 case (panel (e) of Figure \ref{fig:flux_multi_figure}), the material is more neutron
rich, due to the higher peak density and increased time spent at higher densities in this trajectory, favoring electron capture reactions in the accreted material.

\section{Implications for Galactic Chemical Evolution} 
\label{sec:binary_and_GCE}

In the previous section we have explored the large variety of nucleosynthesis that can be found in material infalling close to the neutron star at different accretion rates, and ejected back in the stellar host, and into the interstellar medium following evolution of the companion star. Both light and heavy elements can be made, showing proton-rich or neutron-rich isotopic patterns by slightly changing the trajectory conditions within a realistic parameter space. Interestingly, all of the different nucleosynthesis patterns shown could be produced in the same merging event, during the evolution of the merging system. The toy model presented in section \ref{sec:mdot} made to access the possible nucleosynthesis found in these systems is meant to be a first step to explore the production of elements in these systems, and motivate fully resolved hydrodynamics simulations. Based on the calculations present here, we can now also verify if these systems are relevant also for galactic chemical evolution (GCE).



\subsection{Population synthesis study of CE phases}
\label{sec:binary}

Before studying the role of these stellar objects in a galactic chemical evolution context, we need to determine the yields mass ejection in CE, and couple the accretion rates from Section~\ref{sec:mdot} with the properties of the CE systems.  
To study the population of CE phases, we employ the StarTrack population synthesis code \citep{belczynski2002comprehensive,belczynski2008compact}
to generate a population of binary compact objects
. The code is based on revised formulas from \citet{hurley2000comprehensive}; updated with new wind mass loss prescriptions, calibrated tidal interactions, physical estimation of donor’s binding energy in CE calculations and convection driven, neutrino enhanced supernova engines. A full description of these updates is given in \citet{dominik2012double}. The two most recent updates take into account measurements of initial parameter distributions for massive O stars \citep{sana2012binary}
as well as a correction of a technical bug that has limited the formation of BH-BH binaries for high metallicity (e.g., Z = 0.02).  

With this code, we modeled the binary interactions of 500,000 stars for 3 different metallicities (Table~\ref{tab:popsynth}).  Only a small fraction of the systems actually go through a CE phase.  For each of these systems, we calculate the progenitor star mass, the radius at the onset of the CE phase, and the final separation.  Using the progenitor mass and radius at the onset of the CE phase, we can determine which of our MESA progenitors to use and the time in its evolution.  There is a slight inconsistency 
in our approach, because the formulae for the stellar radii versus time in the population synthesis models are not identical to our MESA models, but this discrepancy allows us to make a first approximation of the accretion rates.  Once we determine the mass and time of the CE interaction, we can use the final separation to determine how deep the neutron star inspirals into its companion.

\begin{table}
	\centering
	\caption{Population Synthesis Calculations}
	\label{tab:popsynth}
	\begin{tabular}{lc} 
		\toprule
		Metallicity & CE systems \\
		\midrule
        0.02 & 7172 \\
        0.002 & 12234 \\
        0.0002 & 7172 \\
	\bottomrule
	\end{tabular}
\end{table}

\begin{figure}	
\includegraphics[width=\columnwidth]{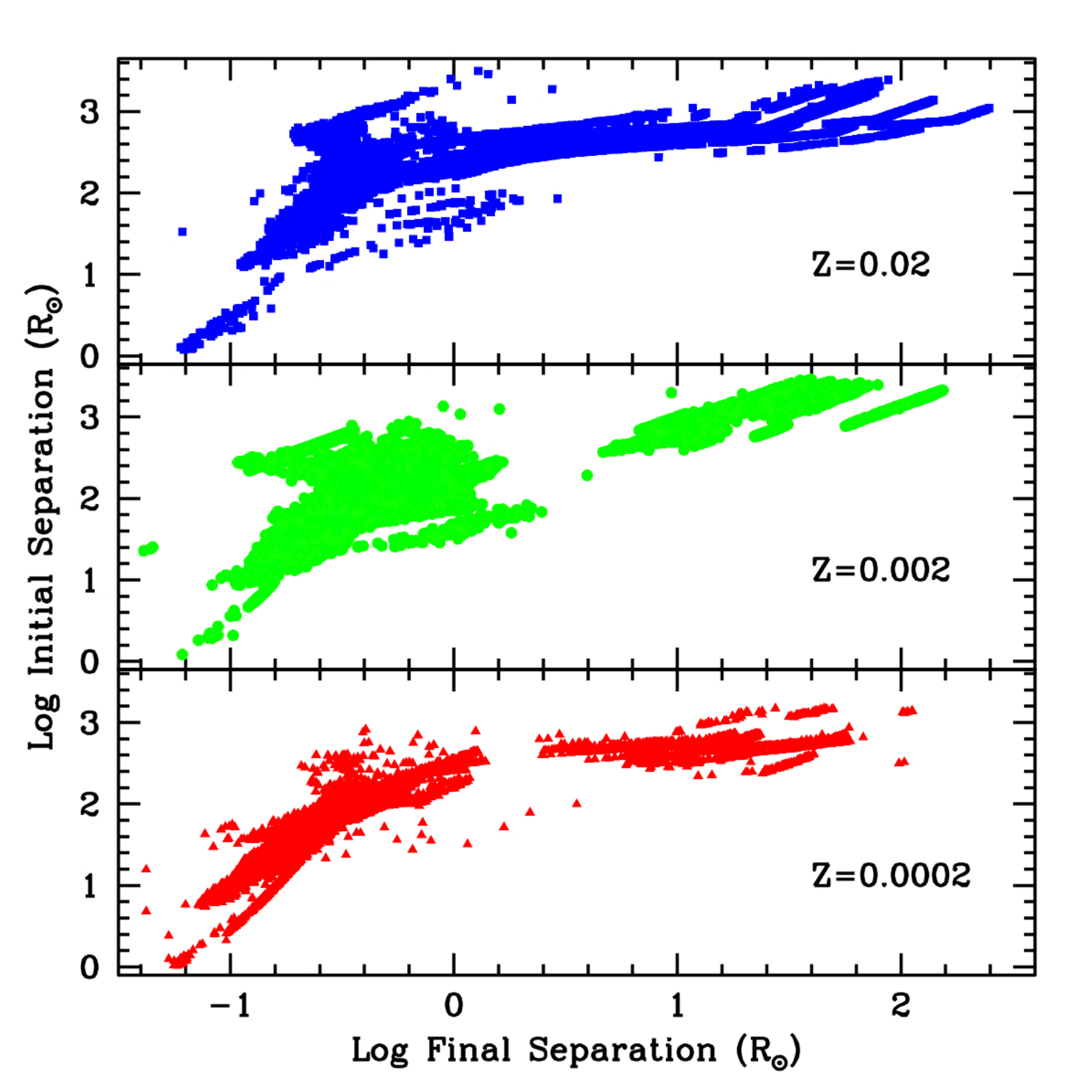}
    \caption{Initial versus final separation for all the binary systems undergoing a CE phase from our population synthesis simulations, given at 3 metallicities.}
    \label{fig:aovaf}
\end{figure}

With our stellar structures and the separation evolution, we can follow the full range of accretion rates in the CE phase.  During the CE phase, the NS is continuously accreting and ejecting mass, evolving through a range of accretion rates as the NS spirals down toward the stellar core.  The accretion and ejection is rapid compared to the CE phase, and we can approximate this evolution as a series of phases with different accretion rates. For each binary system in our population synthesis calculation, we know the mass of the stellar component, the evolutionary timescale of the star at the initiation of CE phase (from the initial binary separation) and the range of accretion rates (based on the final binary separation and our stellar models).  We then assume that 25\% of the accreted mass is reheated and ejected based on the results of accretion simulations\citep{2006ApJ...646L.131F} to get the rate of mass ejection.

To get the total amount of mass ejected, we integrate the mass ejection rate over the inward spiral of the NS.
Based on CE simulations \citep{2008ApJ...672L..41R,2012ApJ...746...74R,2012ApJ...744...52P,2013A&ARv..21...59I}, we assume that the duration of a typical CE phase persists for roughly 3 times the orbital period ($P_{\rm onset}$) at the onset of the CE phase.  In our calculations, we will assume that the time spent at each radius ($t_r$) is 3 times this orbital period ($P_r$) at radius r, for each bin i, normalized by the orbital period of each radius:
\begin{equation}
t_r = 3 P_{\rm onset} P(r_i) / \sum _{i} (P(r_i))
\end{equation}

\begin{figure}
	\includegraphics[width=\columnwidth,clip=true,trim=0cm 3.5cm 0cm 4cm]{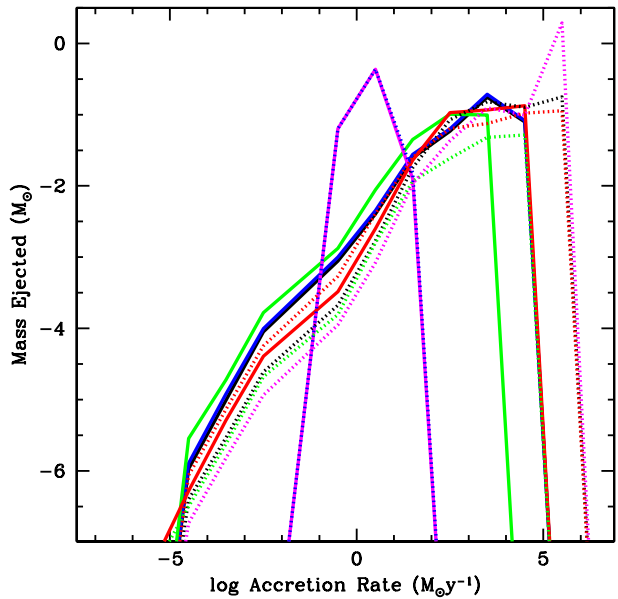}
    \caption{Mass ejected versus accretion rate for 10 sample binary systems in our population synthesis calculations.  As the neutron star spirals into its companion, the accretion rate increases.  Throughout this inspiral phase, mass is ejected and the conditions of this ejecta evolve as the neutron star spirals into deeper and deeper stellar layers.  The peak accretion rate depends upon the structure of the stellar core and the depth of the inspiral.}
    \label{fig:masseji}
\end{figure}

In this paper, we consider 3 options for the mass accretion and ejection:  
$\lambda_{\rm BHL}= 1/4$, 25\% mass ejecta, $\lambda_{\rm BHL}= 1/40$, 25\% mass ejecta, and $\lambda_{\rm BHL}= 1/40$, 10\% mass ejecta. Recall that $\lambda_{\rm BHL}$ is a non-dimensional parameter relating to the efficiency of accretion in our system (see section \ref{sec:mass_accretion_est}).  Figure~\ref{fig:masseji} shows the distribution of accretion rates for 10 of the close binaries in our population synthesis calculation with our model using $\lambda_{\rm BHL}= 1/40$, 25\% mass ejecta.

\begin{figure}
	\includegraphics[width=\columnwidth,clip=true,trim=0cm 3.5cm 0cm 4cm]{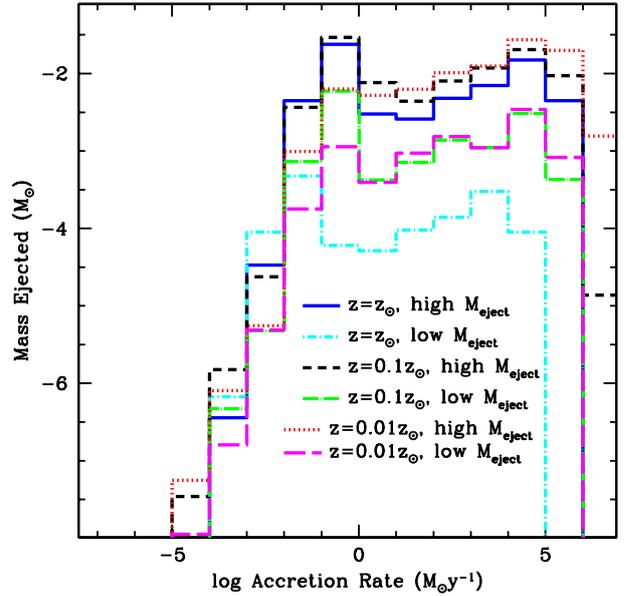}
    \caption{Average mass ejected per binary system (based on the population synthesis models) as a function of metallicity (solar, 1/10th solar and 1/100th solar) for two extremes in our assumptions for accretion rate and mass ejected ("high" denotes $\lambda_{\rm BHL}= 1/4$, 25\% mass ejecta, low denotes $\lambda_{\rm BHL}= 1/40$, 10\% mass ejecta).  Some binaries will not eject much mass at all, others will eject over a solar mass.}
    \label{fig:massej}
\end{figure}

Using our full population of binary systems, we can estimate the average mass ejected per binary system as a function of metallicity and our value for $\lambda_{\rm BHL}$ (Figure~\ref{fig:massej}). The fraction of merging systems in our models are 34\%, 61\% and 75\% for 1, 0.1 and 0.01 solar metalicities repectively. On average, a binary system ejects a few hundredths of a solar mass of material (in practice, some systems can eject nearly a solar mass of material while many systems eject very little mass).

Our MESA stellar models consist of a course grid in mass and time, which presents challenges in the outer layers of the companion star in matching with our population synthesis models, extracted from \citet{hurley2000comprehensive}. This is not a concern however at higher accretion rates near the core of the companion, where the majority of burning in these models is observed, so long as the core masses of our stars are similar. As we focus on higher accretion rates for our yields, inconsistencies introduced by this are minimal in our models.

\subsection{Galactic Chemical Evolution}
\label{sec:gce}
To estimate the contribution of CE mass ejection events in a galactic chemical evolution context, we use the OMEGA code described in \cite{2017ApJ...835..128C}.  This is a classical one-zone open-box model (e.g., \citealt{1980FCPh....5..287T}) that is part of the open-source NuPyCEE package\footnote{\url{https://github.com/NuGrid/NuPyCEE}}.  We adopt the same default Milky Way setup as in \cite{2018ApJ...854..105C}.  Our input parameters are tuned to reproduce the star formation rate, the gas fraction, the gas inflow rate, and the Type~Ia and core-collapse supernova rates currently observed in the Milky Way (see Table~1 in \citealt{2015A&A...580A.126K}), within the observational errors.  Our model is also calibrated to reach solar metallicity  
($Z_{m}=0.014$, \citealt{asplund:09}) when the Sun forms, which is 4.6\,Gyr before the end of the simulation (\citealt{2017GeCoA.201..345C}). The evolution of metallicity in our model is generated using NuGrid yields (\citealt{Ritter2017a}) for massive and low-mass stars, and the yields of \cite{1999ApJS..125..439I} for Type~Ia supernovae.

\begin{figure}
\includegraphics[width=\columnwidth]{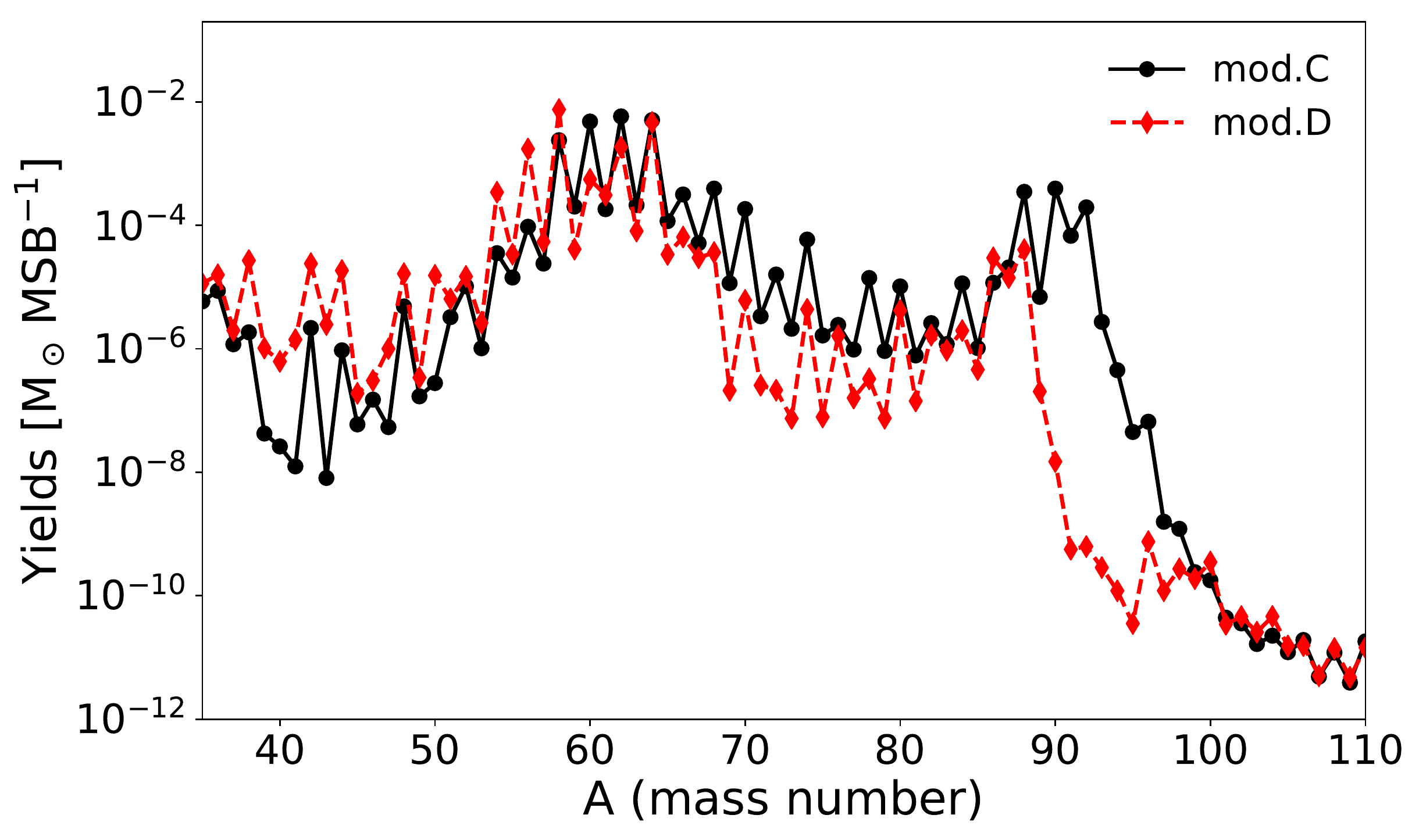}
\caption{Yields ejected by CE events on average per massive star binary (MSB) at $Z=0.02$, representing the convolution of our population synthesis models (Section~\ref{sec:binary}) with the yields calculated for different accretion rates (see Section~\ref{sec: ppn set}), using the trajectories "mod.C" (black line) and "mod.D" (red line). This is for $\lambda_\mathrm{BHL}=1/4$ and 25\% mass ejecta (see Section~\ref{sec:binary}). The complete set of yields used in our galactic chemical evolution calculations for different metallicities and mass accretion and ejection options is available on-line at \href{http://apps.canfar.net/storage/list/nugrid/nb-users/Common_Envelope_Data_Keegans_2018}{CANFAR}.}
\label{fig_yields_GCE}
\end{figure}

To include the contribution of CE events, we use the NuPyCEE delayed-extra source implementation\footnote{\url{https://github.com/NuGrid/NuPyCEE/blob/master/DOC/Capabilities/Delayed_extra_sources.ipynb}} which allows to include additional enrichment sources based on input metallicity-dependent delay-time distribution (DTD) functions and yields.  Because the high NS accretion rate CE events studied here occur in systems involving two massive stars, the mass ejection rate is assumed to follow the lifetime of massive stars.  In practical terms, for each stellar population formed in our model, all CE events occur between 5 and 40~Myr following the formation of the progenitor stars. This first-order implementation will be improved in follow-up studies. For the chemical composition of CE events ejecta, we convolve our nucleosynthesis calculations (see Section~\ref{sec: ppn set}) with the metallicity-dependent mass ejection rates inferred from our population synthesis analysis (Section~\ref{sec:binary}). An example of the resulting yields is shown in Figure~\ref{fig_yields_GCE}. The complete set of yields used in our chemical evolution calculations is available on-line at \href{http://apps.canfar.net/storage/list/nugrid/nb-users/Common_Envelope_Data_Keegans_2018}{CANFAR}.

\begin{figure*}
\includegraphics[width=0.95\textwidth]{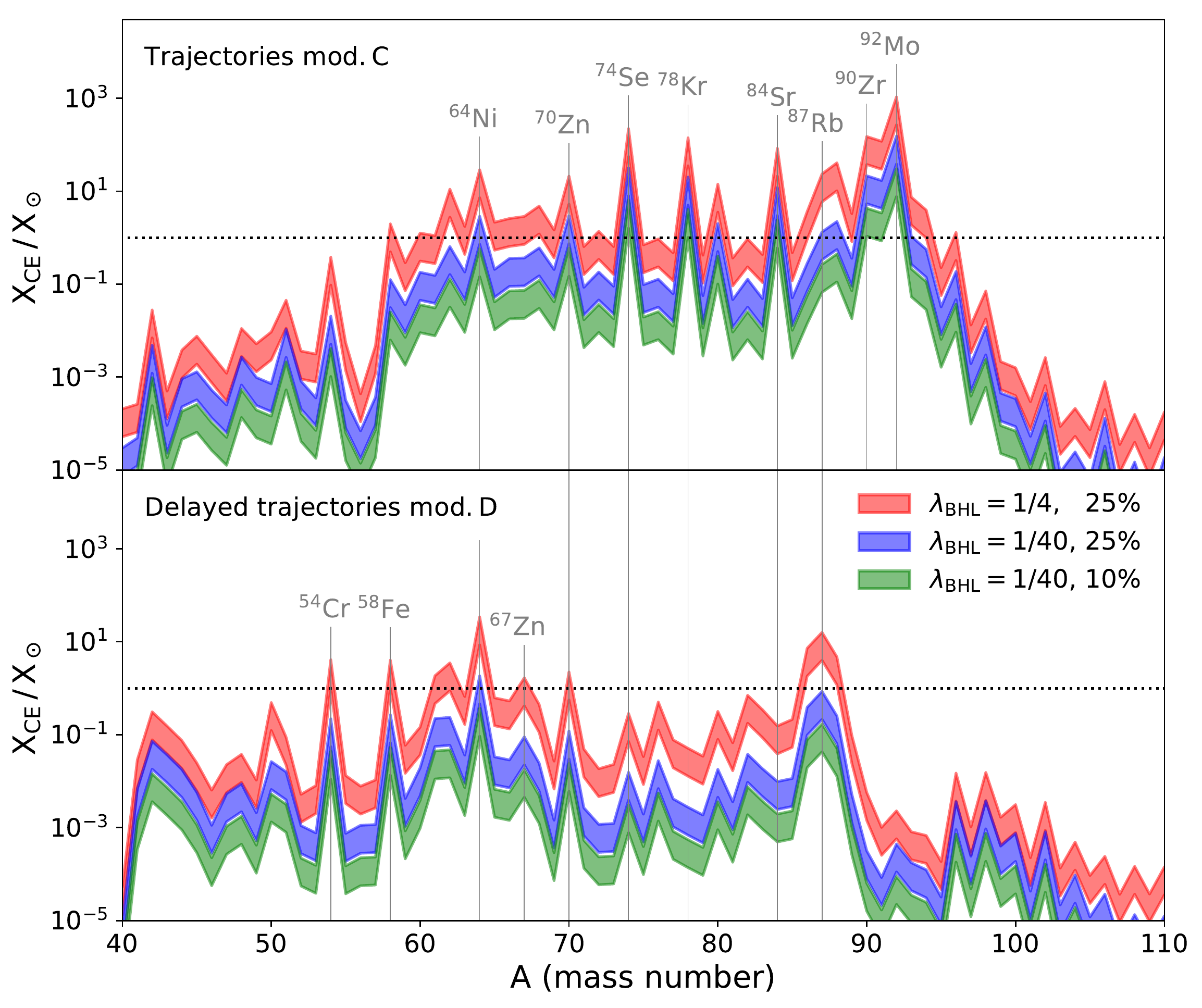}
\caption{Contribution of CE events on the solar isotopic (mass number) composition of the Sun, using the galactic chemical evolution framework described in Section~\ref{sec:gce} and the yields from the trajectories "mod.C" (top) and "mod.D" (bottom).  Different colors represent our different mass accretion and ejection options (see Section~\ref{sec:binary}), where $\lambda_\mathrm{BHL}$ defines the magnitude of the accretion rates (see Equation~\ref{eq_acc_rate}) and the percentage represent the fraction of accreted mass that is reheated and ejected.  For each color, the band represents the range of solutions assuming different binary fractions from 25\,\% to 100\,\% for massive stars.  The dotted horizontal line shows a contribution of 100\,\%.  Anything above this line implies an overestimation relative to the solar composition. Isotopes of interest which are overproduced are labeled.}
\label{fig_GCE_centraj}
\end{figure*}

Figure~\ref{fig_GCE_centraj} shows the contribution of CE events to the solar isotopic composition predicted by our models, using our three mass accretion and ejection options described in Section~\ref{sec:binary}:  $\lambda_{\rm BHL}= 1/4$, 25\% mass ejecta, $\lambda_{\rm BHL}= 1/40$, 25\% mass ejecta, and $\lambda_{\rm BHL}= 1/40$, 10\% mass ejecta. We also varied the binary fraction of massive stars between 25\,\% and 100\,\%.  Overall, the different mass accretion and ejection options cause more variations in our predictions than the binary fraction.  Those figures show that in some cases, CE events could significantly contribute to the chemical evolution of some iron-peak and first-peak neutron-capture isotopes in the Galaxy.  In fact, some isotopes are overproduced in our models relative to the solar composition. Our result here show only the contributions from common envelope events to the solar composition, as direct comparison to isotopic abundances of other contributing events depend sensitively on the models chosen.

As mentioned in the previous sections, the nucleosynthesis of CE events is sensitive to the physical conditions (i.e., temperature and density). For example, when using the yields from the trajectories "mod.C.", the p-isotopes $^{74}$Se, $^{78}$Kr, $^{84}$Sr, and $^{92}$Mo, as well as $^{90,91}$Zr, are always overestimated by more than an order of magnitude. However, when using the yields from the delayed trajectories "mod.D.", none of these isotopes are significantly produced. Instead, the isotopes contributing the most to the solar compositions are rather in that case concentrated on the neutron-rich side, such as $^{64}$Ni, $^{70}$Zn, and $^{86}$Kr. 

We note that the nucleosynthesis has been calculated assuming 
an initial metallicity Z = 0.014, and that we used those yields for all metallicities. The purpose of our chemical evolution calculations is to verify whether or not CE events could be important for galactic chemical evolution. Our results should be seen as a first-order approximation and a motivation for future work.

\section{Final discussion and Conclusions}
\label{sec:conclusions}

In this work we have explored the nucleosynthesis produced by neutron stars accreting in binary common envelopes. A realistic range of accretion rate conditions is explored (between 1$~\rm{{M}_{\odot}yr^{-1}}$ and 10$^5~\rm{{M}_{\odot}yr^{-1}}$), for two sets of trajectories.
In the first set, material is assumed to be near free-fall, and suddenly ejected with acceleration comparable (and in opposite direction) to the gravitational acceleration.
In the second set, material is ejected with more gradual acceleration, resulting in lower temperature and density peaks compared to the first set. 
A large variety of nucleosynthesis patterns were obtained. Heavier elements are produced with increasing accretion rate, due to the higher temperature and density peaks. In particular, weak interactions become extremely important in defining the final composition at accretion rates of 10$^{4}$-10$^{5}~\rm{{M}_{\odot}yr^{-1}}$, and leading to a proton-rich or neutron-rich nucleosynthesis pattern for heavy elements between Fe and Ru. 

We test for the first time the impact of CE events in a galactic chemical evolution context. We find that accreting neutron stars could contribute in a non-negligible way to the solar composition for some isotopes. In particular, using yields from the first set of trajectories, we overproduce many p-isotopes such as $^{74}$Se, $^{78}$Kr, $^{84}$Sr, and $^{92}$Mo, among others. Using the second set of trajectories where there is a gradual change in trajectory, these events do not contribute to the solar abundances of these isotopes. The yields from these events is therefore highly dependent on the specific physical conditions experienced in the CE event, and the conditions in which the nucleosynthesis takes place is therefore crucial for galactic chemical evolution. 

To summarize, we have shown that neutron stars accreting in binary common envelopes are potentially an important (unaccounted for) nucleosynthesis site for the chemical evolution of the Galaxy. Due to the simple approximations made in this first study for the nucleosynthesis trajectories, present sets of yields are still highly uncertain. But these results are a first important step, demanding more detailed simulations in the future.

\section*{Acknowledgments}
NuGrid acknowledges support from NSF grant PHY-1430152 (JINA Center for the Evolution of the Elements) and STFC (through the University of Hull's Consolidated Grant ST/R000840/1). MP and JK also acknowledge support by ongoing resource allocations on the University of Hull's High Performance Computing Facility viper. 

BC acknowledges support from the ERC Consolidator Grant (Hungary) funding scheme (project RADIOSTAR, G.A. n. 724560). 

KB acknowledges support from the Polish National Science Center (NCN) grants
Sonata Bis 2 (DEC-2012/07/E/ST9/01360), OPUS (2015/19/B/ST9/01099), Maestro 2015/18/A/ST9/00746 and LOFT/eXTP 2013/10/M/ST9/00729.

This work was, in part, supported by the US Department of Energy through the Los Alamos National Laboratory. Additional funding was provided by the Laboratory Directed Research and Development Program and the Center for Nonlinear Studies at Los Alamos National Laboratory under project number 20170508DR.

\begin{table*}\label{centraj_abundances_table}
\begin{tabular}{c|c|c|c|c|c|c|c|c|}
\toprule
Element & A & 1$\rm{{M}_{\odot}yr^{-1}}$ & 10$\rm{{M}_{\odot}yr^{-1}}$ & 10$^{2}\rm{{M}_{\odot}yr^{-1}}$ & 10$^{3}\rm{{M}_{\odot}yr^{-1}}$ & 10$^{4}\rm{{M}_{\odot}yr^{-1}}$ & 10$^{5}\rm{{M}_{\odot}yr^{-1}}$\\
\midrule
H          & 1   & 7.27e-01  & 7.27e-01  & 7.27e-01  & 7.24e-01  & 9.71e-12  & 1.27e-21  \\
H          & 2   & 2.20e-18  & 2.39e-18  & 2.60e-18  & 2.81e-18  & 2.39e-25  & 4.72e-34  \\
He         & 3   & 7.20e-12  & 3.44e-15  & 5.93e-15  & 1.16e-14  & 3.42e-29  & 3.15e-24  \\
He         & 4   & 2.59e-01  & 2.54e-01  & 2.23e-01  & 1.44e-01  & 5.03e-02  & 9.54e-06  \\
Li         & 7   & 3.99e-05  & 4.68e-10  & 4.17e-11  & 2.09e-11  & 6.79e-21  & 3.61e-23  \\
B          & 11  & 5.77e-10  & 4.91e-15  & 3.18e-16  & 9.10e-17  & 6.25e-32  & 2.15e-31  \\
C          & 12  & 2.60e-09  & 1.30e-11  & 7.56e-12  & 8.86e-12  & 3.00e-05  & 3.03e-12  \\
C          & 13  & 2.67e-07  & 1.08e-09  & 8.62e-10  & 2.75e-10  & 1.30e-11  & 4.91e-27  \\
N          & 14  & 1.27e-06  & 1.88e-06  & 2.11e-06  & 1.08e-06  & 2.76e-18  & 1.04e-26  \\
N          & 15  & 2.98e-04  & 1.21e-08  & 4.00e-08  & 5.89e-08  & 1.22e-18  & 8.35e-16  \\
O          & 16  & 2.54e-07  & 2.48e-09  & 1.08e-08  & 1.71e-08  & 6.14e-08  & 1.12e-11  \\
O          & 17  & 6.08e-08  & 7.24e-09  & 2.15e-09  & 2.56e-10  & 1.69e-17  & 5.65e-23  \\
O          & 18  & 2.68e-03  & 1.10e-05  & 1.40e-05  & 7.64e-06  & 1.86e-23  & 5.20e-24  \\
F          & 19  & 8.83e-08  & 7.62e-12  & 2.44e-11  & 3.36e-11  & 8.00e-19  & 6.47e-17  \\
Ne         & 20  & 2.57e-05  & 5.33e-08  & 1.05e-07  & 1.01e-07  & 7.16e-08  & 3.38e-13  \\
Ne         & 21  & 3.60e-04  & 1.72e-09  & 2.08e-09  & 5.11e-09  & 4.37e-14  & 6.95e-21  \\
Ne         & 22  & 4.77e-03  & 9.40e-06  & 7.06e-06  & 1.56e-06  & 1.44e-17  & 1.69e-16  \\
Na         & 23  & 3.70e-04  & 9.51e-06  & 1.90e-05  & 1.33e-05  & 1.06e-11  & 4.17e-15  \\
Mg         & 24  & 1.69e-03  & 2.53e-05  & 4.08e-05  & 2.80e-05  & 4.66e-08  & 1.73e-14  \\
Mg         & 25  & 2.05e-05  & 1.77e-07  & 7.95e-07  & 1.46e-06  & 4.15e-13  & 5.87e-16  \\
Mg         & 26  & 8.21e-08  & 1.16e-07  & 2.25e-09  & 4.39e-09  & 2.20e-10  & 2.12e-14  \\
Al         & 27  & 5.67e-04  & 1.34e-04  & 3.88e-05  & 2.06e-05  & 5.77e-08  & 3.16e-14  \\
Si         & 28  & 2.77e-04  & 6.40e-04  & 1.47e-04  & 8.13e-05  & 6.38e-08  & 6.26e-15  \\
Si         & 29  & 1.36e-08  & 4.16e-05  & 5.47e-08  & 6.50e-08  & 3.47e-09  & 8.51e-14  \\
Si         & 30  & 1.79e-06  & 1.02e-06  & 6.21e-09  & 3.47e-09  & 4.32e-07  & 2.49e-13  \\
P          & 31  & 5.47e-04  & 4.39e-03  & 9.98e-05  & 5.76e-05  & 2.16e-08  & 7.18e-15  \\
S          & 32  & 6.98e-05  & 7.98e-03  & 1.50e-04  & 8.73e-05  & 3.05e-07  & 2.40e-13  \\
S          & 33  & 3.07e-04  & 8.05e-04  & 1.45e-06  & 2.21e-06  & 9.65e-08  & 6.35e-14  \\
S          & 34  & 1.17e-08  & 6.46e-05  & 3.48e-07  & 2.64e-08  & 1.30e-06  & 1.19e-12  \\
S          & 36  & 7.00e-99  & 7.00e-99  & 4.63e-35  & 2.73e-35  & 5.05e-10  & 2.51e-13  \\
Cl         & 35  & 1.36e-08  & 1.62e-03  & 1.86e-04  & 1.11e-04  & 2.53e-07  & 3.84e-14  \\
Cl         & 37  & 4.47e-05  & 2.05e-04  & 8.47e-07  & 1.19e-06  & 1.38e-07  & 3.33e-13  \\
Ar         & 36  & 2.38e-06  & 2.43e-03  & 2.56e-04  & 1.57e-04  & 7.30e-07  & 1.40e-14  \\
Ar         & 38  & 8.44e-10  & 1.79e-04  & 1.84e-04  & 1.06e-04  & 4.36e-06  & 3.22e-12  \\
Ar         & 40  & 2.42e-33  & 3.44e-39  & 4.52e-35  & 7.19e-39  & 8.07e-09  & 5.48e-13  \\
K          & 39  & 2.32e-08  & 3.77e-07  & 3.66e-06  & 4.13e-06  & 2.97e-07  & 3.07e-14  \\
K          & 40  & 3.46e-27  & 4.09e-28  & 8.11e-28  & 1.38e-27  & 1.88e-08  & 3.81e-15  \\
K          & 41  & 8.10e-09  & 2.45e-06  & 2.11e-07  & 3.90e-07  & 2.20e-07  & 7.32e-14  \\
Ca         & 40  & 1.09e-10  & 7.40e-08  & 1.10e-06  & 1.17e-06  & 1.34e-06  & 1.18e-16  \\
Ca         & 42  & 5.99e-05  & 5.05e-05  & 1.81e-04  & 7.80e-05  & 9.18e-06  & 9.35e-14  \\
Ca         & 43  & 1.46e-08  & 1.12e-07  & 3.01e-07  & 3.46e-07  & 4.04e-07  & 1.39e-13  \\
Ca         & 44  & 3.65e-07  & 1.37e-05  & 1.01e-04  & 8.75e-05  & 1.27e-06  & 9.41e-13  \\
Ca         & 46  & 9.00e-99  & 9.00e-99  & 9.00e-99  & 5.28e-37  & 7.65e-14  & 2.04e-11  \\
Ca         & 48  & 8.00e-99  & 8.00e-99  & 8.00e-99  & 2.96e-38  & 1.26e-21  & 6.82e-11  \\
Sc         & 45  & 1.16e-08  & 6.51e-08  & 9.97e-07  & 1.96e-06  & 4.01e-06  & 2.08e-13  \\
Ti         & 46  & 1.68e-07  & 8.36e-08  & 1.18e-05  & 9.79e-07  & 8.96e-06  & 3.76e-14  \\
Ti         & 47  & 5.35e-08  & 1.98e-07  & 3.47e-06  & 3.55e-06  & 1.72e-06  & 2.35e-11  \\
Ti         & 48  & 2.32e-08  & 6.27e-07  & 6.72e-04  & 3.89e-04  & 7.55e-06  & 5.11e-09  \\
Ti         & 49  & 7.77e-08  & 7.82e-08  & 5.66e-06  & 8.81e-06  & 8.71e-06  & 1.66e-08  \\
Ti         & 50  & 1.00e-98  & 1.00e-98  & 1.00e-98  & 1.60e-24  & 7.33e-12  & 2.18e-05  \\
V          & 50  & 1.00e-99  & 1.00e-99  & 1.00e-99  & 4.25e-29  & 3.54e-08  & 1.01e-09  \\
$Y_\mathrm{efin}$ & N/A & 8.64e-01  & 8.63e-01  & 8.63e-01  & 8.61e-01  & 4.68e-01  & 4.47e-01  \\
\bottomrule
\end{tabular}
\caption{Abundances for trajectories mod.C.ar1d0, mod.C.ar1d1,
	mod.C.ar1d2, mod.C.ar1d3 mod.C.ar1d4 and mod.C.ar1d5 respectively, up to isotope
	$^{50}$V. Complete tables are available online. Final electron fractions (Ye) are
	provided at the end of the table. 
}
\label{tab:centraj_abunds}
\end{table*}

\begin{table*}
\begin{tabular}{c|c|c|c|c|c|c|c|c|}
\toprule
Element & A & 1$\rm{{M}_{\odot}yr^{-1}}$ & 10$\rm{{M}_{\odot}yr^{-1}}$ & 10$^{2}\rm{{M}_{\odot}yr^{-1}}$ & 10$^{3}\rm{{M}_{\odot}yr^{-1}}$ & 10$^{4}\rm{{M}_{\odot}yr^{-1}}$ & 10$^{5}\rm{{M}_{\odot}yr^{-1}}$\\
\midrule
H          & 1   & 7.28e-01  & 7.27e-01  & 7.26e-01  & 7.25e-01  & 2.79e-01  & 2.92e-23  \\
H          & 2   & 2.20e-18  & 2.39e-18  & 2.60e-18  & 2.81e-18  & 1.07e-18  & 1.13e-34  \\
He         & 3   & 3.49e-11  & 1.02e-14  & 5.71e-15  & 1.12e-14  & 3.65e-15  & 1.58e-23  \\
He         & 4   & 2.61e-01  & 2.58e-01  & 2.42e-01  & 1.57e-01  & 8.04e-02  & 6.70e-07  \\
Li         & 7   & 6.22e-05  & 2.13e-06  & 2.45e-11  & 1.54e-11  & 1.90e-11  & 5.18e-24  \\
B          & 11  & 7.91e-10  & 2.28e-11  & 2.03e-16  & 7.38e-17  & 4.50e-17  & 4.72e-30  \\
C          & 12  & 4.92e-09  & 6.21e-11  & 9.70e-12  & 1.14e-11  & 1.74e-11  & 6.18e-13  \\
C          & 13  & 4.07e-07  & 3.33e-08  & 1.11e-09  & 3.62e-10  & 1.22e-10  & 1.25e-24  \\
N          & 14  & 1.90e-03  & 2.34e-06  & 2.59e-06  & 1.36e-06  & 3.90e-07  & 3.78e-22  \\
N          & 15  & 5.68e-04  & 2.13e-06  & 5.10e-08  & 7.54e-08  & 5.32e-08  & 2.42e-16  \\
O          & 16  & 5.68e-07  & 1.23e-06  & 1.46e-08  & 2.09e-08  & 2.92e-09  & 1.41e-12  \\
O          & 17  & 5.75e-06  & 8.21e-09  & 2.87e-09  & 3.52e-10  & 2.20e-10  & 5.56e-21  \\
O          & 18  & 5.10e-03  & 2.14e-05  & 1.78e-05  & 9.78e-06  & 2.95e-06  & 8.34e-23  \\
F          & 19  & 1.49e-07  & 5.69e-10  & 3.13e-11  & 4.32e-11  & 3.11e-11  & 5.27e-18  \\
Ne         & 20  & 3.91e-07  & 3.31e-05  & 1.43e-07  & 1.24e-07  & 1.66e-08  & 4.38e-15  \\
Ne         & 21  & 2.44e-06  & 4.38e-04  & 2.69e-09  & 6.21e-09  & 1.92e-09  & 3.73e-19  \\
Ne         & 22  & 8.73e-04  & 5.74e-05  & 9.60e-06  & 2.10e-06  & 8.16e-07  & 2.41e-18  \\
Na         & 23  & 5.11e-06  & 2.56e-03  & 3.06e-05  & 1.68e-05  & 4.64e-06  & 8.55e-17  \\
Mg         & 24  & 2.15e-05  & 6.82e-03  & 6.57e-05  & 3.54e-05  & 9.73e-06  & 9.50e-17  \\
Mg         & 25  & 2.58e-06  & 4.82e-04  & 1.30e-06  & 1.85e-06  & 1.20e-06  & 9.90e-18  \\
Mg         & 26  & 5.72e-08  & 1.90e-06  & 3.64e-09  & 5.55e-09  & 3.71e-09  & 2.55e-16  \\
Al         & 27  & 5.22e-04  & 5.60e-04  & 7.50e-05  & 2.68e-05  & 7.86e-06  & 3.12e-16  \\
Si         & 28  & 6.36e-05  & 1.97e-03  & 2.87e-04  & 1.05e-04  & 3.27e-05  & 1.46e-16  \\
Si         & 29  & 2.18e-08  & 6.16e-08  & 1.57e-07  & 8.60e-08  & 7.09e-08  & 4.47e-16  \\
Si         & 30  & 2.07e-06  & 7.29e-07  & 1.11e-08  & 4.61e-09  & 5.70e-09  & 2.22e-14  \\
P          & 31  & 6.08e-04  & 3.01e-04  & 2.69e-04  & 7.81e-05  & 2.20e-05  & 2.76e-16  \\
S          & 32  & 6.52e-06  & 3.34e-04  & 4.06e-04  & 1.18e-04  & 3.41e-05  & 1.61e-15  \\
S          & 33  & 3.06e-04  & 3.07e-04  & 4.59e-06  & 2.99e-06  & 2.01e-06  & 5.10e-15  \\
S          & 34  & 3.97e-09  & 7.29e-08  & 5.65e-06  & 3.99e-08  & 1.81e-08  & 1.72e-13  \\
S          & 36  & 3.98e-31  & 2.21e-34  & 1.75e-35  & 2.49e-35  & 2.45e-35  & 1.39e-12  \\
Cl         & 35  & 8.35e-07  & 5.31e-08  & 2.70e-03  & 1.67e-04  & 4.25e-05  & 8.99e-15  \\
Cl         & 37  & 4.48e-05  & 4.50e-05  & 3.33e-04  & 1.89e-06  & 1.20e-06  & 1.46e-13  \\
Ar         & 36  & 1.47e-06  & 2.49e-06  & 3.72e-03  & 2.36e-04  & 6.26e-05  & 9.50e-17  \\
Ar         & 38  & 1.62e-09  & 2.35e-09  & 6.93e-03  & 1.69e-04  & 4.22e-05  & 3.66e-13  \\
Ar         & 40  & 8.39e-31  & 9.21e-34  & 2.06e-35  & 7.71e-36  & 3.66e-35  & 7.49e-13  \\
K          & 39  & 2.40e-08  & 2.01e-08  & 2.64e-04  & 7.14e-06  & 1.03e-06  & 9.58e-15  \\
K          & 40  & 9.24e-22  & 3.28e-29  & 1.00e-99  & 2.93e-27  & 2.07e-27  & 7.28e-17  \\
K          & 41  & 2.78e-06  & 5.34e-08  & 3.71e-04  & 6.61e-07  & 2.79e-07  & 7.05e-15  \\
Ca         & 40  & 2.22e-05  & 1.12e-10  & 8.46e-05  & 2.05e-06  & 1.45e-07  & 9.98e-20  \\
Ca         & 42  & 3.42e-05  & 5.62e-05  & 5.95e-03  & 1.32e-04  & 5.94e-05  & 6.79e-14  \\
Ca         & 43  & 2.30e-08  & 3.73e-08  & 6.71e-04  & 6.01e-07  & 5.51e-07  & 1.81e-13  \\
Ca         & 44  & 1.67e-07  & 4.06e-06  & 4.76e-03  & 1.48e-04  & 1.20e-05  & 3.79e-11  \\
Ca         & 46  & 4.26e-26  & 9.00e-99  & 9.00e-99  & 6.41e-37  & 8.43e-37  & 1.32e-08  \\
Ca         & 48  & 2.89e-29  & 8.00e-99  & 8.00e-99  & 8.00e-99  & 8.00e-99  & 4.03e-07  \\
Sc         & 45  & 4.49e-08  & 5.72e-08  & 4.29e-05  & 3.32e-06  & 1.05e-06  & 2.25e-11  \\
Ti         & 46  & 6.65e-09  & 5.51e-08  & 2.87e-05  & 1.75e-06  & 1.73e-05  & 5.48e-13  \\
Ti         & 47  & 2.57e-07  & 2.90e-07  & 2.54e-04  & 6.26e-06  & 2.45e-06  & 5.54e-09  \\
Ti         & 48  & 9.20e-07  & 1.68e-07  & 2.89e-03  & 7.22e-04  & 1.42e-04  & 3.65e-08  \\
Ti         & 49  & 4.63e-07  & 7.25e-08  & 3.25e-05  & 1.67e-05  & 7.03e-06  & 4.13e-06  \\
Ti         & 50  & 2.52e-11  & 2.80e-32  & 1.40e-35  & 1.01e-36  & 8.80e-24  & 1.59e-03  \\
V          & 50  & 7.96e-09  & 1.00e-99  & 1.00e-99  & 9.80e-29  & 8.00e-27  & 2.95e-09  \\
$Y_\mathrm{efin}$ & N/A & 8.64e-01  & 8.63e-01  & 8.63e-01  & 8.62e-01  & 6.37e-01  & 4.42e-01  \\
\bottomrule 
  
\end{tabular}
\label{tab:isotopic_yields}
\caption{The same as Table \ref{tab:centraj_abunds}, but for trajectories mod.D.ar1d0,
mod.D.ar1d1, mod.D.ar1d2, mod.D.ar1d3 mod.D.ar1d4 and mod.D.ar1d5 respectively.
}
\label{tab:delayed_abundances_table} 
\end{table*}

\begin{table*}
\begin{tabular}{c|c|c|c|c|c|c|c|c|}
\toprule
Element & A & 1$\rm{{M}_{\odot}yr^{-1}}$ & 10$\rm{{M}_{\odot}yr^{-1}}$ & 10$^{2}\rm{{M}_{\odot}yr^{-1}}$ & 10$^{3}\rm{{M}_{\odot}yr^{-1}}$ & 10$^{4}\rm{{M}_{\odot}yr^{-1}}$ & 10$^{5}\rm{{M}_{\odot}yr^{-1}}$\\
\midrule
H  & 1  & 9.98e-01 & 9.97e-01 & 9.98e-01 & 9.93e-01 & 1.33e-11 & 1.74e-21  \\ 
H  & 2  & 1.56e-13 & 1.69e-13 & 1.84e-13 & 1.99e-13 & 1.69e-20 & 3.34e-29  \\ 
He & 3  & 1.66e-07 & 7.93e-11 & 1.37e-10 & 2.68e-10 & 7.88e-25 & 7.26e-20  \\ 
He & 4  & 9.93e-01 & 9.71e-01 & 8.54e-01 & 5.50e-01 & 1.93e-01 & 3.65e-05  \\ 
Li & 7  & 9.16e+05 & 1.07e+01 & 9.57e-01 & 4.79e-01 & 1.56e-10 & 8.28e-13  \\ 
B  & 11 & 2.17e-01 & 1.85e-06 & 1.20e-07 & 3.43e-08 & 2.36e-23 & 8.12e-23  \\  
C  & 12 & 1.49e-06 & 7.44e-09 & 4.33e-09 & 5.07e-09 & 1.71e-02 & 1.73e-09  \\  
C  & 13 & 1.26e-02 & 5.08e-05 & 4.06e-05 & 1.30e-05 & 6.13e-07 & 2.31e-22  \\  
N  & 14 & 2.51e-03 & 3.73e-03 & 4.18e-03 & 2.15e-03 & 5.47e-15 & 2.06e-23  \\  
N  & 15 & 1.50e+02 & 6.06e-03 & 2.01e-02 & 2.96e-02 & 6.15e-13 & 4.20e-10  \\  
O  & 16 & 5.80e-05 & 5.66e-07 & 2.47e-06 & 3.91e-06 & 1.40e-05 & 2.56e-09  \\  
O  & 17 & 3.51e-02 & 4.17e-03 & 1.24e-03 & 1.48e-04 & 9.75e-12 & 3.26e-17  \\  
O  & 18 & 2.71e+02 & 1.12e+00 & 1.42e+00 & 7.73e-01 & 1.88e-18 & 5.27e-19  \\  
F  & 19 & 2.13e-01 & 1.84e-05 & 5.90e-05 & 8.13e-05 & 1.93e-12 & 1.56e-10  \\  
Ne & 20 & 3.33e-02 & 6.91e-05 & 1.37e-04 & 1.31e-04 & 9.28e-05 & 4.38e-10  \\  
Ne & 21 & 1.85e+02 & 8.84e-04 & 1.07e-03 & 2.63e-03 & 2.25e-08 & 3.58e-15  \\  
Ne & 22 & 7.64e+01 & 1.51e-01 & 1.13e-01 & 2.49e-02 & 2.31e-13 & 2.71e-12  \\  
Na & 23 & 1.81e+01 & 4.66e-01 & 9.31e-01 & 6.50e-01 & 5.21e-07 & 2.04e-10  \\  
Mg & 24 & 4.40e+00 & 6.57e-02 & 1.06e-01 & 7.26e-02 & 1.21e-04 & 4.48e-11  \\  
Mg & 25 & 4.03e-01 & 3.47e-03 & 1.56e-02 & 2.87e-02 & 8.16e-09 & 1.15e-11  \\  
Mg & 26 & 1.41e-03 & 1.99e-03 & 3.86e-05 & 7.54e-05 & 3.77e-06 & 3.65e-10  \\  
Al & 27 & 1.49e+01 & 3.53e+00 & 1.02e+00 & 5.42e-01 & 1.52e-03 & 8.32e-10  \\  
Si & 28 & 5.52e-01 & 1.28e+00 & 2.93e-01 & 1.62e-01 & 1.27e-04 & 1.25e-11  \\  
Si & 29 & 5.14e-04 & 1.58e+00 & 2.07e-03 & 2.46e-03 & 1.31e-04 & 3.23e-09  \\  
Si & 30 & 9.94e-02 & 5.68e-02 & 3.45e-04 & 1.93e-04 & 2.40e-02 & 1.38e-08  \\  
P  & 31 & 1.28e+02 & 1.03e+03 & 2.34e+01 & 1.35e+01 & 5.07e-03 & 1.69e-09  \\  
S  & 32 & 2.77e-01 & 3.17e+01 & 5.95e-01 & 3.47e-01 & 1.21e-03 & 9.53e-10  \\  
S  & 33 & 1.50e+02 & 3.92e+02 & 7.09e-01 & 1.08e+00 & 4.71e-02 & 3.10e-08  \\  
S  & 34 & 9.84e-04 & 5.44e+00 & 2.93e-02 & 2.23e-03 & 1.10e-01 & 9.98e-08  \\  
S  & 36 & 1.38e-91 & 1.38e-91 & 9.14e-28 & 5.38e-28 & 9.96e-03 & 4.94e-06  \\  
Cl & 35 & 2.70e-03 & 3.21e+02 & 3.70e+01 & 2.21e+01 & 5.04e-02 & 7.63e-09  \\  
Cl & 37 & 2.63e+01 & 1.20e+02 & 4.98e-01 & 6.99e-01 & 8.10e-02 & 1.96e-07  \\  
Ar & 36 & 8.61e-02 & 8.80e+01 & 9.26e+00 & 5.66e+00 & 2.64e-02 & 5.06e-10  \\  
Ar & 38 & 1.59e-04 & 3.37e+01 & 3.47e+01 & 2.00e+01 & 8.21e-01 & 6.06e-07  \\  
Ar & 40 & 2.71e-25 & 3.85e-31 & 5.05e-27 & 8.05e-31 & 9.03e-01 & 6.13e-05  \\  
K  & 39 & 8.83e-03 & 1.43e-01 & 1.40e+00 & 1.57e+00 & 1.13e-01 & 1.17e-08  \\  
K  & 40 & 1.03e-17 & 1.21e-18 & 2.41e-18 & 4.09e-18 & 5.58e+01 & 1.13e-05  \\  
K  & 41 & 4.07e-02 & 1.23e+01 & 1.06e+00 & 1.96e+00 & 1.10e+00 & 3.68e-07  \\  
Ca & 40 & 2.30e-06 & 1.56e-03 & 2.32e-02 & 2.46e-02 & 2.83e-02 & 2.49e-12  \\  
Ca & 42 & 1.80e+02 & 1.52e+02 & 5.44e+02 & 2.34e+02 & 2.76e+01 & 2.81e-07  \\  
Ca & 43 & 2.05e-01 & 1.58e+00 & 4.23e+00 & 4.86e+00 & 5.68e+00 & 1.96e-06  \\  
Ca & 44 & 3.24e-01 & 1.22e+01 & 8.99e+01 & 7.78e+01 & 1.13e+00 & 8.37e-07  \\  
Ca & 46 & 3.99e-90 & 3.99e-90 & 3.99e-90 & 2.34e-28 & 3.39e-05 & 9.05e-03  \\  
Ca & 48 & 7.27e-92 & 7.27e-92 & 7.27e-92 & 2.69e-31 & 1.15e-14 & 6.20e-04  \\  
Sc & 45 & 3.82e-01 & 2.15e+00 & 3.29e+01 & 6.48e+01 & 1.32e+02 & 6.87e-06  \\  
Ti & 46 & 9.30e-01 & 4.62e-01 & 6.51e+01 & 5.41e+00 & 4.95e+01 & 2.08e-07  \\  
Ti & 47 & 3.21e-01 & 1.19e+00 & 2.08e+01 & 2.13e+01 & 1.03e+01 & 1.41e-04  \\  
Ti & 48 & 1.37e-02 & 3.72e-01 & 3.99e+02 & 2.31e+02 & 4.48e+00 & 3.03e-03  \\  
Ti & 49 & 6.15e-01 & 6.19e-01 & 4.48e+01 & 6.97e+01 & 6.89e+01 & 1.31e-01  \\  
Ti & 50 & 8.09e-92 & 8.09e-92 & 8.09e-92 & 1.29e-17 & 5.93e-05 & 1.77e+02  \\  
V  & 50 & 1.33e-90 & 1.33e-90 & 1.33e-90 & 5.67e-20 & 4.72e+01 & 1.35e+00  \\
\bottomrule

\end{tabular}
\caption{Overproduction factors for all accretion rates for trajectories mod.C.ar1d0,
	mod.C.ar1d1, mod.C.ar1d2, mod.C.ar1d3 mod.C.ar1d4 and mod.C.ar1d5 respectively, up to
	isotope $^{50}$V. Complete tables are available online. 
    }
\label{tab:overprod_factors_centraj}
\end{table*}

\begin{table*}
\begin{tabular}{c|c|c|c|c|c|c|c|c|}
\toprule
Element & A & 1$\rm{{M}_{\odot}yr^{-1}}$ & 10$\rm{{M}_{\odot}yr^{-1}}$ & 10$^{2}\rm{{M}_{\odot}yr^{-1}}$ & 10$^{3}\rm{{M}_{\odot}yr^{-1}}$ & 10$^{4}\rm{{M}_{\odot}yr^{-1}}$ & 10$^{5}\rm{{M}_{\odot}yr^{-1}}$\\
\midrule
H  & 1  & 9.98e-01 & 9.97e-01 & 9.97e-01 & 9.94e-01 & 3.83e-01 & 4.00e-23  \\ 
H  & 2  & 1.56e-13 & 1.69e-13 & 1.84e-13 & 1.99e-13 & 7.55e-14 & 8.00e-30  \\ 
He & 3  & 8.04e-07 & 2.35e-10 & 1.32e-10 & 2.59e-10 & 8.43e-11 & 3.64e-19  \\ 
He & 4  & 1.00e+00 & 9.89e-01 & 9.28e-01 & 6.03e-01 & 3.08e-01 & 2.56e-06  \\ 
Li & 7  & 1.43e+06 & 4.90e+04 & 5.64e-01 & 3.55e-01 & 4.37e-01 & 1.19e-13  \\ 
B  & 11 & 2.98e-01 & 8.59e-03 & 7.67e-08 & 2.78e-08 & 1.70e-08 & 1.78e-21  \\  
C  & 12 & 2.81e-06 & 3.55e-08 & 5.55e-09 & 6.50e-09 & 9.96e-09 & 3.54e-10  \\  
C  & 13 & 1.92e-02 & 1.57e-03 & 5.21e-05 & 1.71e-05 & 5.75e-06 & 5.88e-20  \\  
N  & 14 & 3.77e+00 & 4.65e-03 & 5.14e-03 & 2.70e-03 & 7.73e-04 & 7.49e-19  \\  
N  & 15 & 2.86e+02 & 1.07e+00 & 2.57e-02 & 3.79e-02 & 2.68e-02 & 1.22e-10  \\  
O  & 16 & 1.30e-04 & 2.80e-04 & 3.34e-06 & 4.77e-06 & 6.68e-07 & 3.23e-10  \\  
O  & 17 & 3.32e+00 & 4.73e-03 & 1.66e-03 & 2.03e-04 & 1.27e-04 & 3.21e-15  \\  
O  & 18 & 5.17e+02 & 2.16e+00 & 1.81e+00 & 9.90e-01 & 2.99e-01 & 8.44e-18  \\  
F  & 19 & 3.59e-01 & 1.37e-03 & 7.56e-05 & 1.04e-04 & 7.50e-05 & 1.27e-11  \\  
Ne & 20 & 5.06e-04 & 4.29e-02 & 1.86e-04 & 1.61e-04 & 2.14e-05 & 5.68e-12  \\  
Ne & 21 & 1.26e+00 & 2.25e+02 & 1.38e-03 & 3.20e-03 & 9.87e-04 & 1.92e-13  \\  
Ne & 22 & 1.40e+01 & 9.19e-01 & 1.54e-01 & 3.37e-02 & 1.31e-02 & 3.87e-14  \\  
Na & 23 & 2.50e-01 & 1.25e+02 & 1.50e+00 & 8.24e-01 & 2.27e-01 & 4.19e-12  \\  
Mg & 24 & 5.58e-02 & 1.77e+01 & 1.70e-01 & 9.19e-02 & 2.52e-02 & 2.46e-13  \\  
Mg & 25 & 5.07e-02 & 9.47e+00 & 2.56e-02 & 3.63e-02 & 2.37e-02 & 1.95e-13  \\  
Mg & 26 & 9.82e-04 & 3.27e-02 & 6.25e-05 & 9.55e-05 & 6.38e-05 & 4.38e-12  \\  
Al & 27 & 1.37e+01 & 1.47e+01 & 1.97e+00 & 7.04e-01 & 2.07e-01 & 8.22e-12  \\  
Si & 28 & 1.27e-01 & 3.94e+00 & 5.73e-01 & 2.10e-01 & 6.51e-02 & 2.92e-13  \\  
Si & 29 & 8.27e-04 & 2.34e-03 & 5.96e-03 & 3.26e-03 & 2.69e-03 & 1.69e-11  \\  
Si & 30 & 1.15e-01 & 4.06e-02 & 6.20e-04 & 2.56e-04 & 3.17e-04 & 1.23e-09  \\  
P  & 31 & 1.43e+02 & 7.07e+01 & 6.32e+01 & 1.83e+01 & 5.16e+00 & 6.48e-11  \\  
S  & 32 & 2.59e-02 & 1.33e+00 & 1.61e+00 & 4.70e-01 & 1.35e-01 & 6.40e-12  \\  
S  & 33 & 1.49e+02 & 1.50e+02 & 2.24e+00 & 1.46e+00 & 9.79e-01 & 2.49e-09  \\  
S  & 34 & 3.35e-04 & 6.14e-03 & 4.76e-01 & 3.36e-03 & 1.52e-03 & 1.45e-08  \\  
S  & 36 & 7.84e-24 & 4.36e-27 & 3.46e-28 & 4.91e-28 & 4.83e-28 & 2.74e-05  \\  
Cl & 35 & 1.66e-01 & 1.05e-02 & 5.37e+02 & 3.33e+01 & 8.44e+00 & 1.79e-09  \\  
Cl & 37 & 2.63e+01 & 2.65e+01 & 1.96e+02 & 1.11e+00 & 7.07e-01 & 8.61e-08  \\  
Ar & 36 & 5.31e-02 & 9.01e-02 & 1.34e+02 & 8.52e+00 & 2.26e+00 & 3.43e-12  \\  
Ar & 38 & 3.05e-04 & 4.42e-04 & 1.30e+03 & 3.18e+01 & 7.95e+00 & 6.90e-08  \\  
Ar & 40 & 9.39e-23 & 1.03e-25 & 2.30e-27 & 8.63e-28 & 4.09e-27 & 8.38e-05  \\  
K  & 39 & 9.15e-03 & 7.66e-03 & 1.01e+02 & 2.72e+00 & 3.94e-01 & 3.65e-09  \\  
K  & 40 & 2.74e-12 & 9.73e-20 & 2.97e-90 & 8.69e-18 & 6.16e-18 & 2.16e-07  \\  
K  & 41 & 1.40e+01 & 2.68e-01 & 1.86e+03 & 3.32e+00 & 1.40e+00 & 3.54e-08  \\  
Ca & 40 & 4.67e-01 & 2.36e-06 & 1.78e+00 & 4.32e-02 & 3.04e-03 & 2.10e-15  \\  
Ca & 42 & 1.03e+02 & 1.69e+02 & 1.79e+04 & 3.96e+02 & 1.79e+02 & 2.04e-07  \\  
Ca & 43 & 3.24e-01 & 5.24e-01 & 9.43e+03 & 8.45e+00 & 7.75e+00 & 2.54e-06  \\  
Ca & 44 & 1.49e-01 & 3.61e+00 & 4.23e+03 & 1.32e+02 & 1.06e+01 & 3.37e-05  \\  
Ca & 46 & 1.89e-17 & 3.99e-90 & 3.99e-90 & 2.84e-28 & 3.74e-28 & 5.85e+00  \\  
Ca & 48 & 2.62e-22 & 7.27e-92 & 7.27e-92 & 7.27e-92 & 7.27e-92 & 3.67e+00  \\  
Sc & 45 & 1.48e+00 & 1.89e+00 & 1.42e+03 & 1.10e+02 & 3.45e+01 & 7.43e-04  \\  
Ti & 46 & 3.68e-02 & 3.04e-01 & 1.59e+02 & 9.69e+00 & 9.56e+01 & 3.03e-06  \\  
Ti & 47 & 1.54e+00 & 1.74e+00 & 1.52e+03 & 3.76e+01 & 1.47e+01 & 3.32e-02  \\  
Ti & 48 & 5.45e-01 & 9.96e-02 & 1.72e+03 & 4.28e+02 & 8.41e+01 & 2.17e-02  \\  
Ti & 49 & 3.67e+00 & 5.74e-01 & 2.57e+02 & 1.32e+02 & 5.57e+01 & 3.26e+01  \\  
Ti & 50 & 2.04e-04 & 2.26e-25 & 1.14e-28 & 8.19e-30 & 7.12e-17 & 1.28e+04  \\  
V  & 50 & 1.06e+01 & 1.33e-90 & 1.33e-90 & 1.31e-19 & 1.07e-17 & 3.93e+00  \\
\bottomrule
\end{tabular}
\caption{The same as Table \ref{tab:overprod_factors_centraj}, but for trajectories
mod.D.ar1d0, mod.D.ar1d1, mod.D.ar1d2, mod.D.ar1d3 mod.D.ar1d4 and mod.D.ar1d5 respectively.
}
\label{tab:overprod_factors_delay}
\end{table*}




\bibliographystyle{mnras}
\bibliography{master}

\begin{thebibliography}{}
\makeatletter
\relax
\def\mn@urlcharsother{\let\do\@makeother \do\$\do\&\do\#\do\^\do\_\do\%\do\~}
\def\mn@doi{\begingroup\mn@urlcharsother \@ifnextchar [ {\mn@doi@}
  {\mn@doi@[]}}
\def\mn@doi@[#1]#2{\def\@tempa{#1}\ifx\@tempa\@empty \href
  {http://dx.doi.org/#2} {doi:#2}\else \href {http://dx.doi.org/#2} {#1}\fi
  \endgroup}
\def\mn@eprint#1#2{\mn@eprint@#1:#2::\@nil}
\def\mn@eprint@arXiv#1{\href {http://arxiv.org/abs/#1} {{\tt arXiv:#1}}}
\def\mn@eprint@dblp#1{\href {http://dblp.uni-trier.de/rec/bibtex/#1.xml}
  {dblp:#1}}
\def\mn@eprint@#1:#2:#3:#4\@nil{\def\@tempa {#1}\def\@tempb {#2}\def\@tempc
  {#3}\ifx \@tempc \@empty \let \@tempc \@tempb \let \@tempb \@tempa \fi \ifx
  \@tempb \@empty \def\@tempb {arXiv}\fi \@ifundefined
  {mn@eprint@\@tempb}{\@tempb:\@tempc}{\expandafter \expandafter \csname
  mn@eprint@\@tempb\endcsname \expandafter{\@tempc}}}

\bibitem[\protect\citeauthoryear{{Abbott} et~al.,}{{Abbott}
  et~al.}{2017}]{2017ApJ...848L..12A}
{Abbott} B.~P.,  et~al., 2017, \mn@doi [\apjl] {10.3847/2041-8213/aa91c9},
  \href {http://adsabs.harvard.edu/abs/2017ApJ...848L..12A} {848, L12}

\bibitem[\protect\citeauthoryear{{Angulo}, {Arnould}  \& {Rayet, M. et
  al.}}{{Angulo} et~al.}{1999}]{angulo:99}
{Angulo} C.,  {Arnould} M.,   {Rayet, M. et al.} 1999, A 656, 3

\bibitem[\protect\citeauthoryear{{Arcones} \& {Montes}}{{Arcones} \&
  {Montes}}{2011}]{arcones:11}
{Arcones} A.,  {Montes} F.,  2011, \mn@doi [\apj] {10.1088/0004-637X/731/1/5},
  \href {http://adsabs.harvard.edu/abs/2011ApJ...731....5A} {731, 5}

\bibitem[\protect\citeauthoryear{{Arnould} \& {Goriely}}{{Arnould} \&
  {Goriely}}{2003}]{arnould:03}
{Arnould} M.,  {Goriely} S.,  2003, \mn@doi [\physrep]
  {10.1016/S0370-1573(03)00242-4}, \href
  {http://adsabs.harvard.edu/abs/2003PhR...384....1A} {384, 1}

\bibitem[\protect\citeauthoryear{{Asplund}, {Grevesse}, {Sauval}  \&
  {Scott}}{{Asplund} et~al.}{2009}]{asplund:09}
{Asplund} M.,  {Grevesse} N.,  {Sauval} A.~J.,   {Scott} P.,  2009, \mn@doi
  [ARA\&A] {10.1146/annurev.astro.46.060407.145222}, \href
  {http://adsabs.harvard.edu/abs/2009ARA%26A..47..481A} {47, 481}

\bibitem[\protect\citeauthoryear{Belczynski, Kalogera  \& Bulik}{Belczynski
  et~al.}{2002}]{belczynski2002comprehensive}
Belczynski K.,  Kalogera V.,   Bulik T.,  2002, The Astrophysical Journal, 572,
  407

\bibitem[\protect\citeauthoryear{Belczynski, Kalogera, Rasio, Taam, Zezas,
  Bulik, Maccarone  \& Ivanova}{Belczynski
  et~al.}{2008}]{belczynski2008compact}
Belczynski K.,  Kalogera V.,  Rasio F.~A.,  Taam R.~E.,  Zezas A.,  Bulik T.,
  Maccarone T.~J.,   Ivanova N.,  2008, The Astrophysical Journal Supplement
  Series, 174, 223

\bibitem[\protect\citeauthoryear{{Bloom}, {Sigurdsson}  \& {Pols}}{{Bloom}
  et~al.}{1999}]{1999MNRAS.305..763B}
{Bloom} J.~S.,  {Sigurdsson} S.,   {Pols} O.~R.,  1999, \mn@doi [\mnras]
  {10.1046/j.1365-8711.1999.02437.x}, \href
  {http://adsabs.harvard.edu/abs/1999MNRAS.305..763B} {305, 763}

\bibitem[\protect\citeauthoryear{{Bondi}}{{Bondi}}{1952}]{1952MNRAS.112..195B}
{Bondi} H.,  1952, \mn@doi [\mnras] {10.1093/mnras/112.2.195}, \href
  {http://adsabs.harvard.edu/abs/1952MNRAS.112..195B} {112, 195}

\bibitem[\protect\citeauthoryear{Caughlan \& Fowler}{Caughlan \&
  Fowler}{1988}]{caughlan1988thermonuclear}
Caughlan G.~R.,  Fowler W.~A.,  1988, Atomic Data and Nuclear Data Tables, 40,
  283

\bibitem[\protect\citeauthoryear{{Champagne} \& {Wiescher}}{{Champagne} \&
  {Wiescher}}{1992}]{champagne:92}
{Champagne} A.~E.,  {Wiescher} M.,  1992, \mn@doi [Annual Review of Nuclear and
  Particle Science] {10.1146/annurev.ns.42.120192.000351}, \href
  {http://adsabs.harvard.edu/abs/1992ARNPS..42...39C} {42, 39}

\bibitem[\protect\citeauthoryear{{Chevalier}}{{Chevalier}}{1989}]{1989ApJ...346..847C}
{Chevalier} R.~A.,  1989, \mn@doi [\apj] {10.1086/168066}, \href
  {http://adsabs.harvard.edu/abs/1989ApJ...346..847C} {346, 847}

\bibitem[\protect\citeauthoryear{{Chevalier}}{{Chevalier}}{1993}]{1993ApJ...411L..33C}
{Chevalier} R.~A.,  1993, \mn@doi [\apjl] {10.1086/186905}, \href
  {http://adsabs.harvard.edu/abs/1993ApJ...411L..33C} {411, L33}

\bibitem[\protect\citeauthoryear{{Connelly}, {Bollard}  \&
  {Bizzarro}}{{Connelly} et~al.}{2017}]{2017GeCoA.201..345C}
{Connelly} J.~N.,  {Bollard} J.,   {Bizzarro} M.,  2017, \mn@doi [\gca]
  {10.1016/j.gca.2016.10.044}, \href
  {http://adsabs.harvard.edu/abs/2017GeCoA.201..345C} {201, 345}

\bibitem[\protect\citeauthoryear{{C{\^o}t{\'e}}, {O'Shea}, {Ritter}, {Herwig}
  \& {Venn}}{{C{\^o}t{\'e}} et~al.}{2017}]{2017ApJ...835..128C}
{C{\^o}t{\'e}} B.,  {O'Shea} B.~W.,  {Ritter} C.,  {Herwig} F.,   {Venn} K.~A.,
   2017, \mn@doi [\apj] {10.3847/1538-4357/835/2/128}, \href
  {http://adsabs.harvard.edu/abs/2017ApJ...835..128C} {835, 128}

\bibitem[\protect\citeauthoryear{{C{\^o}t{\'e}}, {Denissenkov}, {Herwig},
  {Ruiter}, {Ritter}, {Pignatari}  \& {Belczynski}}{{C{\^o}t{\'e}}
  et~al.}{2018}]{2018ApJ...854..105C}
{C{\^o}t{\'e}} B.,  {Denissenkov} P.,  {Herwig} F.,  {Ruiter} A.~J.,  {Ritter}
  C.,  {Pignatari} M.,   {Belczynski} K.,  2018, \mn@doi [\apj]
  {10.3847/1538-4357/aaaae8}, \href
  {http://adsabs.harvard.edu/abs/2018ApJ...854..105C} {854, 105}

\bibitem[\protect\citeauthoryear{{Cox}, {Vauclair}  \& {Zahn}}{{Cox}
  et~al.}{1983a}]{1983apum.conf.....C}
{Cox} A.~N.,  {Vauclair} S.,   {Zahn} J.~P.,  eds, 1983a, {Astrophysical
  process in upper main sequence stars}

\bibitem[\protect\citeauthoryear{{Cox}, {Vauclair}  \& {Zahn}}{{Cox}
  et~al.}{1983b}]{cox83}
{Cox} A.~N.,  {Vauclair} S.,   {Zahn} J.~P.,  eds, 1983b, {Astrophysical
  process in upper main sequence stars}

\bibitem[\protect\citeauthoryear{{Cyburt}}{{Cyburt}}{2011}]{cyburt:11}
{Cyburt} R.,  2011, in APS Division of Nuclear Physics Meeting Abstracts. p.~7

\bibitem[\protect\citeauthoryear{{Dillmann}, {Heil}, {K{\"a}ppeler}, {Plag},
  {Rauscher}  \& {Thielemann}}{{Dillmann} et~al.}{2006}]{dillmann:06}
{Dillmann} I.,  {Heil} M.,  {K{\"a}ppeler} F.,  {Plag} R.,  {Rauscher} T.,
  {Thielemann} F.-K.,  2006, in {Woehr} A.,  {Aprahamian} A.,  eds,  American
  Institute of Physics Conference Series Vol. 819, Capture Gamma-Ray
  Spectroscopy and Related Topics. pp 123--127, \mn@doi{10.1063/1.2187846}

\bibitem[\protect\citeauthoryear{{Dominik}, {Belczynski}, {Fryer}, {Holz},
  {Berti}, {Bulik}, {Mandel}  \& {O'Shaughnessy}}{{Dominik}
  et~al.}{2012a}]{2012ApJ...759...52D}
{Dominik} M.,  {Belczynski} K.,  {Fryer} C.,  {Holz} D.~E.,  {Berti} E.,
  {Bulik} T.,  {Mandel} I.,   {O'Shaughnessy} R.,  2012a, \mn@doi [\apj]
  {10.1088/0004-637X/759/1/52}, \href
  {http://adsabs.harvard.edu/abs/2012ApJ...759...52D} {759, 52}

\bibitem[\protect\citeauthoryear{Dominik, Belczynski, Fryer, Holz, Berti,
  Bulik, Mandel  \& O'Shaughnessy}{Dominik et~al.}{2012b}]{dominik2012double}
Dominik M.,  Belczynski K.,  Fryer C.,  Holz D.~E.,  Berti E.,  Bulik T.,
  Mandel I.,   O'Shaughnessy R.,  2012b, The Astrophysical Journal, 759, 52

\bibitem[\protect\citeauthoryear{{Dominik}, {Belczynski}, {Fryer}, {Holz},
  {Berti}, {Bulik}, {Mandel}  \& {O'Shaughnessy}}{{Dominik}
  et~al.}{2013}]{2013ApJ...779...72D}
{Dominik} M.,  {Belczynski} K.,  {Fryer} C.,  {Holz} D.~E.,  {Berti} E.,
  {Bulik} T.,  {Mandel} I.,   {O'Shaughnessy} R.,  2013, \mn@doi [\apj]
  {10.1088/0004-637X/779/1/72}, \href
  {http://adsabs.harvard.edu/abs/2013ApJ...779...72D} {779, 72}

\bibitem[\protect\citeauthoryear{{Dominik} et~al.,}{{Dominik}
  et~al.}{2015}]{2015ApJ...806..263D}
{Dominik} M.,  et~al., 2015, \mn@doi [\apj] {10.1088/0004-637X/806/2/263},
  \href {http://adsabs.harvard.edu/abs/2015ApJ...806..263D} {806, 263}

\bibitem[\protect\citeauthoryear{{Fr{\"o}hlich}, {Mart{\'{\i}}nez-Pinedo},
  {Liebend{\"o}rfer}, {Thielemann}, {Bravo}, {Hix}, {Langanke}  \&
  {Zinner}}{{Fr{\"o}hlich} et~al.}{2006}]{froehlich:06}
{Fr{\"o}hlich} C.,  {Mart{\'{\i}}nez-Pinedo} G.,  {Liebend{\"o}rfer} M.,
  {Thielemann} F.-K.,  {Bravo} E.,  {Hix} W.~R.,  {Langanke} K.,   {Zinner}
  N.~T.,  2006, \mn@doi [Physical Review Letters]
  {10.1103/PhysRevLett.96.142502}, \href
  {http://adsabs.harvard.edu/abs/2006PhRvL..96n2502F} {96, 142502}

\bibitem[\protect\citeauthoryear{{Fryer}}{{Fryer}}{2009}]{2009ApJ...699..409F}
{Fryer} C.~L.,  2009, \mn@doi [\apj] {10.1088/0004-637X/699/1/409}, \href
  {http://adsabs.harvard.edu/abs/2009ApJ...699..409F} {699, 409}

\bibitem[\protect\citeauthoryear{{Fryer} \& {Woosley}}{{Fryer} \&
  {Woosley}}{1998}]{1998ApJ...502L...9F}
{Fryer} C.~L.,  {Woosley} S.~E.,  1998, \mn@doi [\apjl] {10.1086/311493}, \href
  {http://adsabs.harvard.edu/abs/1998ApJ...502L...9F} {502, L9}

\bibitem[\protect\citeauthoryear{{Fryer} \& {Young}}{{Fryer} \&
  {Young}}{2007}]{fryer07}
{Fryer} C.~L.,  {Young} P.~A.,  2007, \mn@doi [\apj] {10.1086/513003}, \href
  {http://adsabs.harvard.edu/abs/2007ApJ...659.1438F} {659, 1438}

\bibitem[\protect\citeauthoryear{{Fryer}, {Benz}  \& {Herant}}{{Fryer}
  et~al.}{1996}]{1996ApJ...460..801F}
{Fryer} C.~L.,  {Benz} W.,   {Herant} M.,  1996, \mn@doi [\apj]
  {10.1086/177011}, \href {http://adsabs.harvard.edu/abs/1996ApJ...460..801F}
  {460, 801}

\bibitem[\protect\citeauthoryear{{Fryer}, {Colgate}  \& {Pinto}}{{Fryer}
  et~al.}{1999a}]{1999ApJ...511..885F}
{Fryer} C.~L.,  {Colgate} S.~A.,   {Pinto} P.~A.,  1999a, \mn@doi [\apj]
  {10.1086/306701}, \href {http://adsabs.harvard.edu/abs/1999ApJ...511..885F}
  {511, 885}

\bibitem[\protect\citeauthoryear{{Fryer}, {Woosley}  \& {Hartmann}}{{Fryer}
  et~al.}{1999b}]{1999ApJ...526..152F}
{Fryer} C.~L.,  {Woosley} S.~E.,   {Hartmann} D.~H.,  1999b, \mn@doi [\apj]
  {10.1086/307992}, \href {http://adsabs.harvard.edu/abs/1999ApJ...526..152F}
  {526, 152}

\bibitem[\protect\citeauthoryear{{Fryer}, {Herwig}, {Hungerford}  \&
  {Timmes}}{{Fryer} et~al.}{2006}]{2006ApJ...646L.131F}
{Fryer} C.~L.,  {Herwig} F.,  {Hungerford} A.,   {Timmes} F.~X.,  2006, \mn@doi
  [\apjl] {10.1086/507071}, \href
  {http://adsabs.harvard.edu/abs/2006ApJ...646L.131F} {646, L131}

\bibitem[\protect\citeauthoryear{{Fryer}, {Belczynski}, {Berger}, {Th{\"o}ne},
  {Ellinger}  \& {Bulik}}{{Fryer} et~al.}{2013}]{2013ApJ...764..181F}
{Fryer} C.~L.,  {Belczynski} K.,  {Berger} E.,  {Th{\"o}ne} C.,  {Ellinger} C.,
    {Bulik} T.,  2013, \mn@doi [\apj] {10.1088/0004-637X/764/2/181}, \href
  {http://adsabs.harvard.edu/abs/2013ApJ...764..181F} {764, 181}

\bibitem[\protect\citeauthoryear{{Fuller}, {Fowler}  \& {Newman}}{{Fuller}
  et~al.}{1985}]{fuller:85}
{Fuller} G.~M.,  {Fowler} W.~A.,   {Newman} M.~J.,  1985, \mn@doi [ApJ]
  {10.1086/163208}, \href {http://adsabs.harvard.edu/abs/1985ApJ...293....1F}
  {293, 1}

\bibitem[\protect\citeauthoryear{{Fynbo} et~al.,}{{Fynbo}
  et~al.}{2005}]{fynbo:05}
{Fynbo} H.~O.~U.,  et~al., 2005, Nature, \href
  {http://adsabs.harvard.edu/cgi-bin/nph-bib_query?bibcode=2005Natur.433..136F&db_key=AST}
  {433, 136}

\bibitem[\protect\citeauthoryear{{Goriely}}{{Goriely}}{1999}]{goriely:99}
{Goriely} S.,  1999, \aap, \href
  {http://adsabs.harvard.edu/abs/1999A%26A...342..881G} {342, 881}

\bibitem[\protect\citeauthoryear{{Heil} et~al.,}{{Heil} et~al.}{2008}]{heil:08}
{Heil} M.,  et~al., 2008, \mn@doi [\prc] {10.1103/PhysRevC.78.025803}, \href
  {http://adsabs.harvard.edu/abs/2008PhRvC..78b5803H} {78, 025803}

\bibitem[\protect\citeauthoryear{{Hoyle} \& {Lyttleton}}{{Hoyle} \&
  {Lyttleton}}{1941}]{1941MNRAS.101..227H}
{Hoyle} F.,  {Lyttleton} R.~A.,  1941, \mn@doi [\mnras]
  {10.1093/mnras/101.4.227}, \href
  {http://adsabs.harvard.edu/abs/1941MNRAS.101..227H} {101, 227}

\bibitem[\protect\citeauthoryear{Hurley, Pols  \& Tout}{Hurley
  et~al.}{2000}]{hurley2000comprehensive}
Hurley J.~R.,  Pols O.~R.,   Tout C.~A.,  2000, Monthly Notices of the Royal
  Astronomical Society, 315, 543

\bibitem[\protect\citeauthoryear{Iliadis, D’Auria, Starrfield, Thompson  \&
  Wiescher}{Iliadis et~al.}{2001}]{iliadis2001proton}
Iliadis C.,  D’Auria J.~M.,  Starrfield S.,  Thompson W.~J.,   Wiescher M.,
  2001, The Astrophysical Journal Supplement Series, 134, 151

\bibitem[\protect\citeauthoryear{{Imbriani} et~al.,}{{Imbriani}
  et~al.}{2005}]{imbriani:05}
{Imbriani} G.,  et~al., 2005, \mn@doi [European Physical Journal A]
  {10.1140/epja/i2005-10138-7}, \href
  {http://adsabs.harvard.edu/abs/2005EPJA...25..455I} {25, 455}

\bibitem[\protect\citeauthoryear{{Ivanova} et~al.,}{{Ivanova}
  et~al.}{2013}]{2013A&ARv..21...59I}
{Ivanova} N.,  et~al., 2013, \mn@doi [\aapr] {10.1007/s00159-013-0059-2}, \href
  {http://adsabs.harvard.edu/abs/2013A%26ARv..21...59I} {21, 59}

\bibitem[\protect\citeauthoryear{{Iwamoto}, {Brachwitz}, {Nomoto}, {Kishimoto},
  {Umeda}, {Hix}  \& {Thielemann}}{{Iwamoto}
  et~al.}{1999}]{1999ApJS..125..439I}
{Iwamoto} K.,  {Brachwitz} F.,  {Nomoto} K.,  {Kishimoto} N.,  {Umeda} H.,
  {Hix} W.~R.,   {Thielemann} F.-K.,  1999, \mn@doi [\apjs] {10.1086/313278},
  \href {http://adsabs.harvard.edu/abs/1999ApJS..125..439I} {125, 439}

\bibitem[\protect\citeauthoryear{{Jaeger}, {Kunz}, {Mayer}, {Hammer}, {Staudt},
  {Kratz}  \& {Pfeiffer}}{{Jaeger} et~al.}{2001}]{jaeger:01}
{Jaeger} M.,  {Kunz} R.,  {Mayer} A.,  {Hammer} J.~W.,  {Staudt} G.,  {Kratz}
  K.~L.,   {Pfeiffer} B.,  2001, Physical Review Letters, \href
  {http://adsabs.harvard.edu/abs/2001PhRvL..87t2501J} {87, 202501}

\bibitem[\protect\citeauthoryear{{Jones} et~al.,}{{Jones}
  et~al.}{2013}]{Jones2013a}
{Jones} S.,  et~al., 2013, \mn@doi [\apj] {10.1088/0004-637X/772/2/150}, \href
  {http://adsabs.harvard.edu/abs/2013ApJ...772..150J} {772, 150}

\bibitem[\protect\citeauthoryear{{Kiminki} \& {Kobulnicky}}{{Kiminki} \&
  {Kobulnicky}}{2012}]{2012ApJ...751....4K}
{Kiminki} D.~C.,  {Kobulnicky} H.~A.,  2012, \mn@doi [\apj]
  {10.1088/0004-637X/751/1/4}, \href
  {http://adsabs.harvard.edu/abs/2012ApJ...751....4K} {751, 4}

\bibitem[\protect\citeauthoryear{{Kiminki} et~al.,}{{Kiminki}
  et~al.}{2007}]{2007ApJ...664.1102K}
{Kiminki} D.~C.,  et~al., 2007, \mn@doi [\apj] {10.1086/513709}, \href
  {http://adsabs.harvard.edu/abs/2007ApJ...664.1102K} {664, 1102}

\bibitem[\protect\citeauthoryear{{Kiminki}, {Kobulnicky}, {Gilbert}, {Bird}  \&
  {Chunev}}{{Kiminki} et~al.}{2009}]{2009AJ....137.4608K}
{Kiminki} D.~C.,  {Kobulnicky} H.~A.,  {Gilbert} I.,  {Bird} S.,   {Chunev} G.,
   2009, \mn@doi [\aj] {10.1088/0004-6256/137/6/4608}, \href
  {http://adsabs.harvard.edu/abs/2009AJ....137.4608K} {137, 4608}

\bibitem[\protect\citeauthoryear{{Kobulnicky} \& {Fryer}}{{Kobulnicky} \&
  {Fryer}}{2007}]{2007ApJ...670..747K}
{Kobulnicky} H.~A.,  {Fryer} C.~L.,  2007, \mn@doi [\apj] {10.1086/522073},
  \href {http://adsabs.harvard.edu/abs/2007ApJ...670..747K} {670, 747}

\bibitem[\protect\citeauthoryear{{Kobulnicky} et~al.,}{{Kobulnicky}
  et~al.}{2014}]{2014ApJS..213...34K}
{Kobulnicky} H.~A.,  et~al., 2014, \mn@doi [\apjs]
  {10.1088/0067-0049/213/2/34}, \href
  {http://adsabs.harvard.edu/abs/2014ApJS..213...34K} {213, 34}

\bibitem[\protect\citeauthoryear{{Kratz}, {Bitouzet}, {Thielemann}, {Moeller}
  \& {Pfeiffer}}{{Kratz} et~al.}{1993}]{kratz:93}
{Kratz} K.-L.,  {Bitouzet} J.-P.,  {Thielemann} F.-K.,  {Moeller} P.,
  {Pfeiffer} B.,  1993, \mn@doi [\apj] {10.1086/172196}, \href
  {http://adsabs.harvard.edu/abs/1993ApJ...403..216K} {403, 216}

\bibitem[\protect\citeauthoryear{{Kubryk}, {Prantzos}  \&
  {Athanassoula}}{{Kubryk} et~al.}{2015}]{2015A&A...580A.126K}
{Kubryk} M.,  {Prantzos} N.,   {Athanassoula} E.,  2015, \mn@doi [\aap]
  {10.1051/0004-6361/201424171}, \href
  {http://adsabs.harvard.edu/abs/2015A%26A...580A.126K} {580, A126}

\bibitem[\protect\citeauthoryear{{Kunz}, {Fey}, {Jaeger}, {Mayer}, {Hammer},
  {Staudt}, {Harissopulos}  \& {Paradellis}}{{Kunz} et~al.}{2002}]{kunz:02}
{Kunz} R.,  {Fey} M.,  {Jaeger} M.,  {Mayer} A.,  {Hammer} J.~W.,  {Staudt} G.,
   {Harissopulos} S.,   {Paradellis} T.,  2002, \mn@doi [\apj]
  {10.1086/338384}, \href {http://adsabs.harvard.edu/abs/2002ApJ...567..643K}
  {567, 643}

\bibitem[\protect\citeauthoryear{{Langanke} \&
  {Mart{\'{\i}}nez-Pinedo}}{{Langanke} \&
  {Mart{\'{\i}}nez-Pinedo}}{2000}]{langanke:00}
{Langanke} K.,  {Mart{\'{\i}}nez-Pinedo} G.,  2000, \mn@doi [Nuclear Physics A]
  {10.1016/S0375-9474(00)00131-7}, \href
  {http://adsabs.harvard.edu/abs/2000NuPhA.673..481L} {673, 481}

\bibitem[\protect\citeauthoryear{MacLeod \& Ramirez-Ruiz}{MacLeod \&
  Ramirez-Ruiz}{2014}]{macleod2014accretion}
MacLeod M.,  Ramirez-Ruiz E.,  2014, The Astrophysical Journal Letters, 798,
  L19

\bibitem[\protect\citeauthoryear{{MacLeod} \& {Ramirez-Ruiz}}{{MacLeod} \&
  {Ramirez-Ruiz}}{2015}]{2015APS..APR.U2004M}
{MacLeod} M.,  {Ramirez-Ruiz} E.,  2015, in APS April Meeting Abstracts. p.
  U2.004

\bibitem[\protect\citeauthoryear{{MacLeod}, {Antoni}, {Murguia-Berthier},
  {Macias}  \& {Ramirez-Ruiz}}{{MacLeod} et~al.}{2017}]{2017ApJ...838...56M}
{MacLeod} M.,  {Antoni} A.,  {Murguia-Berthier} A.,  {Macias} P.,
  {Ramirez-Ruiz} E.,  2017, \mn@doi [\apj] {10.3847/1538-4357/aa6117}, \href
  {http://adsabs.harvard.edu/abs/2017ApJ...838...56M} {838, 56}

\bibitem[\protect\citeauthoryear{{Murguia-Berthier}, {MacLeod}, {Ramirez-Ruiz},
  {Antoni}  \& {Macias}}{{Murguia-Berthier} et~al.}{2017}]{2017ApJ...845..173M}
{Murguia-Berthier} A.,  {MacLeod} M.,  {Ramirez-Ruiz} E.,  {Antoni} A.,
  {Macias} P.,  2017, \mn@doi [\apj] {10.3847/1538-4357/aa8140}, \href
  {http://adsabs.harvard.edu/abs/2017ApJ...845..173M} {845, 173}

\bibitem[\protect\citeauthoryear{{Oda}, {Hino}, {Muto}, {Takahara}  \&
  {Sato}}{{Oda} et~al.}{1994}]{oda:94}
{Oda} T.,  {Hino} M.,  {Muto} K.,  {Takahara} M.,   {Sato} K.,  1994, Atomic
  Data and Nuclear Data Tables, \href
  {http://adsabs.harvard.edu/abs/1994ADNDT..56..231O} {56, 231}

\bibitem[\protect\citeauthoryear{{Passy} et~al.,}{{Passy}
  et~al.}{2012}]{2012ApJ...744...52P}
{Passy} J.-C.,  et~al., 2012, \mn@doi [\apj] {10.1088/0004-637X/744/1/52},
  \href {http://adsabs.harvard.edu/abs/2012ApJ...744...52P} {744, 52}

\bibitem[\protect\citeauthoryear{{Paxton}, {Bildsten}, {Dotter}, {Herwig},
  {Lesaffre}  \& {Timmes}}{{Paxton} et~al.}{2011}]{Paxton2011a}
{Paxton} B.,  {Bildsten} L.,  {Dotter} A.,  {Herwig} F.,  {Lesaffre} P.,
  {Timmes} F.,  2011, \mn@doi [\apjs] {10.1088/0067-0049/192/1/3}, \href
  {http://adsabs.harvard.edu/abs/2011ApJS..192....3P} {192, 3}

\bibitem[\protect\citeauthoryear{{Paxton} et~al.,}{{Paxton}
  et~al.}{2013}]{Paxton2013a}
{Paxton} B.,  et~al., 2013, \mn@doi [\apjs] {10.1088/0067-0049/208/1/4}, \href
  {http://adsabs.harvard.edu/abs/2013ApJS..208....4P} {208, 4}

\bibitem[\protect\citeauthoryear{{Paxton} et~al.,}{{Paxton}
  et~al.}{2015}]{Paxton2015a}
{Paxton} B.,  et~al., 2015, \mn@doi [\apjs] {10.1088/0067-0049/220/1/15}, \href
  {http://adsabs.harvard.edu/abs/2015ApJS..220...15P} {220, 15}

\bibitem[\protect\citeauthoryear{{Paxton} et~al.,}{{Paxton}
  et~al.}{2018}]{Paxton2018a}
{Paxton} B.,  et~al., 2018, \mn@doi [\apjs] {10.3847/1538-4365/aaa5a8}, \href
  {http://adsabs.harvard.edu/abs/2018ApJS..234...34P} {234, 34}

\bibitem[\protect\citeauthoryear{{Pignatari} et~al.,}{{Pignatari}
  et~al.}{2016}]{pignatari:16}
{Pignatari} M.,  et~al., 2016, \mn@doi [\apjs] {10.3847/0067-0049/225/2/24},
  \href {http://adsabs.harvard.edu/abs/2016ApJS..225...24P} {225, 24}

\bibitem[\protect\citeauthoryear{{Popham}, {Woosley}  \& {Fryer}}{{Popham}
  et~al.}{1999}]{1999ApJ...518..356P}
{Popham} R.,  {Woosley} S.~E.,   {Fryer} C.,  1999, \mn@doi [\apj]
  {10.1086/307259}, \href {http://adsabs.harvard.edu/abs/1999ApJ...518..356P}
  {518, 356}

\bibitem[\protect\citeauthoryear{{Rauscher}, {Dauphas}, {Dillmann},
  {Fr{\"o}hlich}, {F{\"u}l{\"o}p}  \& {Gy{\"u}rky}}{{Rauscher}
  et~al.}{2013}]{rauscher:13}
{Rauscher} T.,  {Dauphas} N.,  {Dillmann} I.,  {Fr{\"o}hlich} C.,
  {F{\"u}l{\"o}p} Z.,   {Gy{\"u}rky} G.,  2013, \mn@doi [Reports on Progress in
  Physics] {10.1088/0034-4885/76/6/066201}, \href
  {http://adsabs.harvard.edu/abs/2013RPPh...76f6201R} {76, 066201}

\bibitem[\protect\citeauthoryear{{Ricker} \& {Taam}}{{Ricker} \&
  {Taam}}{2008}]{2008ApJ...672L..41R}
{Ricker} P.~M.,  {Taam} R.~E.,  2008, \mn@doi [\apjl] {10.1086/526343}, \href
  {http://adsabs.harvard.edu/abs/2008ApJ...672L..41R} {672, L41}

\bibitem[\protect\citeauthoryear{{Ricker} \& {Taam}}{{Ricker} \&
  {Taam}}{2012}]{2012ApJ...746...74R}
{Ricker} P.~M.,  {Taam} R.~E.,  2012, \mn@doi [\apj]
  {10.1088/0004-637X/746/1/74}, \href
  {http://adsabs.harvard.edu/abs/2012ApJ...746...74R} {746, 74}

\bibitem[\protect\citeauthoryear{{Ritter}, {Herwig}, {Jones}, {Pignatari},
  {Fryer}  \& {Hirschi}}{{Ritter} et~al.}{2017}]{Ritter2017a}
{Ritter} C.,  {Herwig} F.,  {Jones} S.,  {Pignatari} M.,  {Fryer} C.,
  {Hirschi} R.,  2017, preprint, \href
  {http://adsabs.harvard.edu/abs/2017arXiv170908677R} {} (\mn@eprint {arXiv}
  {1709.08677})

\bibitem[\protect\citeauthoryear{Ritter, Herwig, Jones, Pignatari, Fryer  \&
  Hirschi}{Ritter et~al.}{2018}]{ritter2018nugrid}
Ritter C.,  Herwig F.,  Jones S.,  Pignatari M.,  Fryer C.,   Hirschi R.,
  2018, Monthly Notices of the Royal Astronomical Society, 480, 538

\bibitem[\protect\citeauthoryear{{Roberts}, {Woosley}  \& {Hoffman}}{{Roberts}
  et~al.}{2010}]{roberts:10}
{Roberts} L.~F.,  {Woosley} S.~E.,   {Hoffman} R.~D.,  2010, \mn@doi [\apj]
  {10.1088/0004-637X/722/1/954}, \href
  {http://adsabs.harvard.edu/abs/2010ApJ...722..954R} {722, 954}

\bibitem[\protect\citeauthoryear{{Ruffert}}{{Ruffert}}{1994a}]{1994A&AS..106..505R}
{Ruffert} M.,  1994a, \aaps, \href
  {http://adsabs.harvard.edu/abs/1994A%26AS..106..505R} {106, 505}

\bibitem[\protect\citeauthoryear{{Ruffert}}{{Ruffert}}{1994b}]{1994ApJ...427..342R}
{Ruffert} M.,  1994b, \mn@doi [\apj] {10.1086/174144}, \href
  {http://adsabs.harvard.edu/abs/1994ApJ...427..342R} {427, 342}

\bibitem[\protect\citeauthoryear{{Ruffert} \& {Arnett}}{{Ruffert} \&
  {Arnett}}{1994}]{1994ApJ...427..351R}
{Ruffert} M.,  {Arnett} D.,  1994, \mn@doi [\apj] {10.1086/174145}, \href
  {http://adsabs.harvard.edu/abs/1994ApJ...427..351R} {427, 351}

\bibitem[\protect\citeauthoryear{{Sana} et~al.,}{{Sana}
  et~al.}{2012a}]{2012Sci...337..444S}
{Sana} H.,  et~al., 2012a, \mn@doi [Science] {10.1126/science.1223344}, \href
  {http://adsabs.harvard.edu/abs/2012Sci...337..444S} {337, 444}

\bibitem[\protect\citeauthoryear{Sana et~al.,}{Sana
  et~al.}{2012b}]{sana2012binary}
Sana H.,  et~al., 2012b, Science, 337, 444

\bibitem[\protect\citeauthoryear{{Schatz} et~al.,}{{Schatz}
  et~al.}{2001}]{schatz:01}
{Schatz} H.,  et~al., 2001, \mn@doi [Physical Review Letters]
  {10.1103/PhysRevLett.86.3471}, \href
  {http://adsabs.harvard.edu/abs/2001PhRvL..86.3471S} {86, 3471}

\bibitem[\protect\citeauthoryear{{Seeger}, {Fowler}  \& {Clayton}}{{Seeger}
  et~al.}{1965}]{seeger:65}
{Seeger} P.~A.,  {Fowler} W.~A.,   {Clayton} D.~D.,  1965, \mn@doi [\apjs]
  {10.1086/190111}, \href {http://adsabs.harvard.edu/abs/1965ApJS...11..121S}
  {11, 121}

\bibitem[\protect\citeauthoryear{Siegel, Barnes  \& Metzger}{Siegel
  et~al.}{2018}]{siegel2018neutron}
Siegel D.~M.,  Barnes J.,   Metzger B.~D.,  2018, arXiv preprint
  arXiv:1810.00098

\bibitem[\protect\citeauthoryear{{Soker} \& {Gilkis}}{{Soker} \&
  {Gilkis}}{2018}]{2018MNRAS.475.1198S}
{Soker} N.,  {Gilkis} A.,  2018, \mn@doi [\mnras] {10.1093/mnras/stx3287},
  \href {http://adsabs.harvard.edu/abs/2018MNRAS.475.1198S} {475, 1198}

\bibitem[\protect\citeauthoryear{{Thorne} \& {Zytkow}}{{Thorne} \&
  {Zytkow}}{1975}]{1975ApJ...199L..19T}
{Thorne} K.~S.,  {Zytkow} A.~N.,  1975, \mn@doi [\apjl] {10.1086/181839}, \href
  {http://adsabs.harvard.edu/abs/1975ApJ...199L..19T} {199, L19}

\bibitem[\protect\citeauthoryear{{Tinsley}}{{Tinsley}}{1980}]{1980FCPh....5..287T}
{Tinsley} B.~M.,  1980, \fcp, \href
  {http://adsabs.harvard.edu/abs/1980FCPh....5..287T} {5, 287}

\bibitem[\protect\citeauthoryear{{Wanajo}}{{Wanajo}}{2013}]{wanajo:13}
{Wanajo} S.,  2013, \mn@doi [\apjl] {10.1088/2041-8205/770/2/L22}, \href
  {http://adsabs.harvard.edu/abs/2013ApJ...770L..22W} {770, L22}

\bibitem[\protect\citeauthoryear{{Zhang} \& {Fryer}}{{Zhang} \&
  {Fryer}}{2001}]{2001ApJ...550..357Z}
{Zhang} W.,  {Fryer} C.~L.,  2001, \mn@doi [\apj] {10.1086/319734}, \href
  {http://adsabs.harvard.edu/abs/2001ApJ...550..357Z} {550, 357}

\makeatother
\end{thebibliography}



\appendix(Abundances for selected Isotopes) \label{Abundances_appendix}


\bsp	
\label{lastpage}
\end{document}